\documentclass[fleqn, usenatbib]{mnras}

\usepackage{newtxtext, newtxmath}
\usepackage[T1]{fontenc}

\DeclareRobustCommand{\VAN}[3]{#2}
\let\VANthebibliography\thebibliography
\def\thebibliography{%
  \DeclareRobustCommand{\VAN}[3]{##3}\VANthebibliography}

\usepackage{graphicx}
\usepackage{multirow}

\newcommand{\vb}{\textcolor{blue}{Viswanathan, Bystr{\"o}m et al.}}

\title[Early-stripped Sagittarius dwarf galaxy stars]{%
  Tracing the very early disruption of the Sagittarius dwarf
  galaxy in the distant Milky Way halo%
  \thanks{This paper includes data gathered with the 6.5 meter Magellan 
          Telescopes located at Las Campanas Observatory, Chile.}
}

\author[M. Bayer et al.]{%
  Manuel Bayer,$^1$\thanks{E-mail: mbayer@astro.rug.nl}
  Else Starkenburg,$^1$
  Akshara Viswanathan,$^2$
  Vedant Chandra,$^{3,4}$
  Alexander P. Ji$^{5,6,7}$\newauthor
  and Guillaume F. Thomas$^{8,9}$
  \\
  $^1$Kapteyn Astronomical Institute, University of Groningen,
  Landleven 12, NL-9747AD Groningen, the Netherlands\\
  $^2$Dept. of Physics and Astronomy, University of 
  Victoria, P.O. Box 3055, STN CSC, Victoria BC V8W 3P6, Canada\\
  $^3$Center for Astrophysics | Harvard \& Smithsonian, 60 
  Garden St, Cambridge, MA 02138, USA\\
  $^4$Max-Planck-Institut f{\"u}r Astronomie, Königstuhl 17, D-69117 
  Heidelberg, Germany\\
  $^5$Department of Astronomy \& Astrophysics, University of Chicago, 
  5640 South Ellis Avenue, Chicago, IL 60637, USA\\
  $^6$Kavli Institute for Cosmological Physics, University of Chicago, 
  Chicago, IL 60637, USA\\
  $^7$NSF-Simons AI Institute for the Sky (SkAI), 172 E. Chestnut St., 
  Chicago, IL 60611, USA\\
  $^8$Instituto de Astrof{\'i}sica de Canarias, E-38205 La Laguna, 
  Tenerife, Spain\\
  $^9$Universidad de La Laguna, Dpto. Astrof{\'i}sica, E-38206 La 
  Laguna, Tenerife, Spain}

\date{Accepted XXX. Received YYY; in original form ZZZ}

\pubyear{\the\year{}}

\begin{document}
\label{firstpage}
\pagerange{\pageref{firstpage}--\pageref{lastpage}}
\maketitle

\begin{abstract}
  Current models predict that at distances beyond 80~kpc in the
  Milky Way halo, we can find the earliest escaped stars from the
  merging Sagittarius dwarf galaxy. However, observational data on
  the Sagittarius stream at these distances is limited. This study
  examines an overdensity of red giant branch (RGB) stars
  potentially linked to
  Sagittarius merger debris. Using the Magellan Inamori Kyocera
  Echelle spectrograph of Las Campanas Observatory's Clay Telescope,
  we measured the radial velocities and metallicities of these
  stars. We compared their properties with model predictions of
  Sagittarius' disruption and other stellar tracers from the Dark
  Energy Spectroscopic Instrument Data Release 1 and RR Lyrae
  catalogs. Our spectral analysis confirms the significant tight
  clustering of four of these RGBs in full 6D phase space. This
  tight clump is embedded within a larger spur-like feature of the
  Sagittarius stream in the southern sky. A comparison with
  Sagittarius stream models further strengthens this hypothesis and
  shows that this far spur could be composed of stars originally in
  the halo of the Sagittarius dwarf galaxy, stripped in the earliest
  phases of the interaction. The metallicity dispersion of the four
  stars of $0.15 ^ {+0.17} _ {-0.08}$ around the average of
  [Fe/H] = $-1.46 ^ {+0.11} _ {-0.09}$ is very low. This study
  provides the first spectroscopic view of the distant southern
  spur of Sagittarius, composed of stars likely stripped from
  Sagittarius's halo.
\end{abstract}


\section{Introduction}

The stellar stream formed by the infall of the Sagittarius dwarf galaxy
onto the Milky Way is wrapping around the Galaxy
\cite[e.\,g.,][]{ibata_et_al2001, majewski_et_al2003, helmi2004,
                 johnston_et_al2005, belokurov_et_al2006,
                 belokurov_et_al2014, sesar_et_al2017b,
                 antoja_et_al2020, ibata_et_al2020, ramos_et_al2020,
                 ramos_et_al2022}
covering a significant distance range, with the most distant reported
features going out to 140~kpc
\citep{sesar_et_al2017b, e_starkenburg_et_al2019,
       manuel_bayer_et_al2025}.
Detailed modeling of these observed major tidal features can help us to
constrain the Milky Way potential at large distances, understand the
merger process, and reconstruct the progenitor of the Sagittarius dwarf
galaxy
\cite[e.\,g.,][]{fellhauer_et_al2006, koposov_et_al2010,
                 david_r_law_majewski2010, dierickx_loeb2017,
                 thomas_et_al2017, fardal_et_al2019,
                 vasiliev_et_al2021, oria_et_al2022,
                 limberg_et_al2023,
                 vasiliev_et_al2021, oria_et_al2022,
                 elliot_y_davies_et_al2024}.
Such models show that, typically, stars lost from the progenitor of the
Sagittarius dwarf galaxy in the earliest phases of the infall reside
at some of the most distant parts of the Sagittarius stream
\cite[e.g.,][]{david_r_law_majewski2010, fardal_et_al2019,
               vasiliev_et_al2021}.
                                            
The presence or absence of distant spur- or plume-like
features in the Sagittarius stream may moreover be dependent on the
nature of dark matter \cite[e.g.,][]{hainje_et_al_arxiv2503_15589}
and/or theory of gravity
\cite[see, e.g.,]%
[for simulations of the disruption of the Sagittarius dwarf galaxy
 based on modified Newtonian dynamics]{thomas_et_al2017}.
Recently, \citet{hainje_et_al_arxiv2503_15589} suggested that
self-interacting dark matter could amplify the mass loss of the
progenitor of the Sagittarius dwarf galaxy through scattering of dark
matter particles in the progenitor and the Milky Way. An increased mass
loss leaves less mass in the progenitor to protect the stars during the
pericentric passages. The strength of spur- and plume-like features in
the Sagittarius stream are dependent on the mass of the progenitor
during the previous pericentric passage. Therefore, less remaining mass
from the previous pericentric passage leads to weakended spur- and
plume-like features.

Observationally, these distant regions can only be probed with luminous
globular clusters and stars, such as red giant branch stars (RGBs), blue
horizontal branch stars (BHBs), or RR Lyrae
(e.g., \citealt{massari_et_al2017ngc2419}; \citealt{sesar_et_al2017b};
 \citealt{sangmo_tony_sohn_et_al2018};
 \citealt{e_starkenburg_et_al2019};
 \citealt{bellazzini_et_al2020globular_clusters};
 \vb\ \citeyear{viswanathan_bystroem_et_al_arxiv2408_17250};
 \citealt{manuel_bayer_et_al2025}; \citealt{muraveva_et_al2025};
 \citealt{vedant_chandra_et_al2026}).
The spur feature in the trailing stream at $\Lambda = 172^\circ$ in the
heliocentric Sagittarius stream coordinate system%
\footnote{We use the heliocentric, right-handed, spherical coordinate
          system defined by the orbit of the Sagittarius dwarf galaxy,
          as described in \citet{vasiliev_et_al2021} and implemented in
          \texttt{Gala} \citep{gala}. We denote the longitude-like
          angle in this coordinate system by $\Lambda$ and the
          latitude-like angle by $B$.}
identified initially by \citet{sesar_et_al2017b} using variable RR 
Lyrae stars and also observed later with blue horizontal branch stars
\citep{guillaume_f_thomas_et_al2018, e_starkenburg_et_al2019,
       manuel_bayer_et_al2025}
is currently the most distant extension known of this stream with
associated stars extending to heliocentric distances of up to
140~kpc. Interestingly, in the figure 1 of
\citet{sesar_et_al2017b} the mapped Sagittarius stream with
(candidate) RR Lyrae stars also indicate a plume of stars between 
$270^\circ < \Lambda < 315^\circ$ that could go as far out in the halo
as 120~kpc \cite[see also][]{hernitschek_et_al2017}. The on-sky
location of this feature, around (R.A., Dec.) =
($340^{\circ}$, $-12^{\circ}$), coincides with a spur-like feature that
is predicted in the simulations of the infall of the Sagittarius dwarf
galaxy onto the Galaxy by \citet{dierickx_loeb2017} and
\citet{vasiliev_et_al2021}. According to \citet{vasiliev_et_al2021} the
stars in this spur-like feature in the southern sky may probe the
earliest phase of the gravitational interaction between the progenitor
of the Sagittarius dwarf galaxy and the Milky Way.
There are notable differences in the predictions for the distant,
northern, and southern spurs in the simulations by
\citet{dierickx_loeb2017}, \citet{vasiliev_et_al2021}, and
\citet{oria_et_al2022}. These differences include the timing of
particle loss, density, line-of-sight velocities, and heliocentric
distance. High-quality observations are helpful to constrain the models
in these regimes.

In this work, we spectroscopically uncover the southern Sagittarius
spur, and thereby shed light on the earliest phases of the infall of
the Sagittarius dwarf galaxy, through a study of bright RGB stars.
Utilizing parallax information and infrared colors,
\citet{vedant_chandra_et_al2023b} created a catalog of RGB stars to
study halo substructures at large distances. Here, we present the
results of this catalog on the Sagittarius stream spur in the southern
sky, including dedicated spectroscopic follow-up of several stars that
showed up as a significant overdensity in phase space in this region.

The paper is structured as follows: first, we will outline the discovery
of a tight, comoving cluster of RGB stars and describe their
spectroscopic follow-up in \autoref{sec_data}. We detail the data
analysis for these stars in \autoref{sec_analysis}. In
\autoref{sec_results}, we first assess the significance of this clump of
RGB stars, then establish its connection with the Sagittarius
southern spur feature, embedding these four stars into the broader
Sagittarius stream and providing a first overview on this southern spur
feature. A detailed analysis of the differences of the northern and
southern spur feature is also presented in the results section. In
\autoref{sec_discussion}, we discuss the results and the possibility
that the tight cluster of four stars could represent debris from a
disrupted Sagittarius globular cluster. We also briefly discuss future
prospects, before turning to the conclusions in
\autoref{sec_conclusion}.

\section{Data}
\label{sec_data}

\subsection{Target Selection}\label{target_selection_subsection}

We select an all-sky sample of RGB stars using the parallax and
infrared color selection described in
\citet{vedant_chandra_et_al2023b} and \citet{Chandra2025a}
\citep[see also][]{majewski_et_al2003, Conroy2018, Conroy2021}.
Briefly, \textit{Gaia} parallaxes are used to remove obvious
foreground dwarfs, and \textit{WISE} colors are used to distinguish
luminous RGB stars from the far more numerous foreground dwarfs.
These WISE colors cuts exploit pressure-sensitive absorption
differences between dwarf and giant stars, enabling the efficient
selection of cool red giant stars
\citep[e.g.,][]{majewski_et_al2003, Conroy2018}. Approximate
distances to these candidate giant stars are estimated using MIST
isochrones, assuming a metallicity of [Fe/H]$= -1.5$ and an age of
$10$~Gyr. This sample has already been used to target spectroscopic
surveys of the outer halo, and has been found to contain
$\lesssim 10\%$ contamination from foreground dwarf stars
\citep{Chandra2025a, vedant_chandra_et_al2026}.
A key goal of the sample was to search for low surface-brightness
substructures like dwarf galaxies and stellar streams. Such
structures are challenging to detect in current photometric surveys,
but identifying clusters of distant RGB stars selected with high
confidence can help reveal the underlying population of stars and
limit false detections caused by the clustering of different types
of stars \citep[e.g.,][]{Torrealba2019, Chandra2022, Aganze2025}.

On the sample of RGB stars with isochrone-estimated distances
$> 50$~kpc, we run a simple search for pairs of stars that appear
close on the sky, have similar estimated distances, and have similar
proper motions. A similar search technique was performed on H3
Survey data to discover the faint, currently disrupting, `Specter'
dwarf galaxy \citep{Chandra2022}. Specifically, for each star in the
sample, we calculate the 2D vector difference in proper motions for
all neighbors within $1^\circ$ and with consistent distances at the
$2\sigma$ level. We then select `comoving' pairs as those with
proper motions within $2\sigma$ of each other, taking into account
the full covariance matrix from \textit{Gaia}. This type of search
reveals hundreds of pairs in known dwarf galaxies and stellar
streams. After masking out known dwarf galaxies, we are left with a
handful of clusters that do not obviously belong to known streams.

One such cluster lies in the constellation of Aquarius and was
initially suspected to be a new low surface brightness dwarf galaxy
satellite, although it was quickly realised that their positions and
motions followed the predictions of the Sagittarius stream from the
simulation by \citet{vasiliev_et_al2021}. The cluster consists of a
very close pair of stars within $0.1^\circ$ of one another, and
another three giants within $2^\circ$ with similar proper motions.
These stars have magnitudes $16.9\leq G\leq 17.7$, making them
convenient for spectroscopic follow-up observations. We provide a
summary of these five stars in \autoref{tab:targets}%
\footnote{All photometry presented and used in this paper, includes
          dust reddening corrections based on the integrated 2D
          E($B\! -\!V$) reddening map by \citet{sfd} renormalized by
          \citet{schlafly_finkbeiner2011}.}.

\begin{table*}
  \caption{Targets. The table provides the \textit{Gaia} (Early) Data
           Release \citep{gaia_mission, gedr3, gdr3} source IDs,
           equatorial coordinates, parallaxes, and proper motions from
           the \textit{Gaia} (Early) Data Release 3 catalog
           \citep{lindegren_et_al2021}, and de-reddened Pan-STARRS
           magnitudes.}
  \begin{tabular}{lllllllllll}
    \hline
    Source ID & R.A. & Dec. & \multicolumn{2}{l}{$\varpi$}
    & \multicolumn{2}{l}{$\mu ^* _ \alpha$}
    & \multicolumn{2}{l}{$\mu _ \delta$}
    & \multicolumn{2}{l}{$g _ {\text{PS},0}$} \\
    & \multicolumn{2}{l}{J2016} & \multicolumn{2}{l}{[mas]}
    & \multicolumn{2}{l}{$\left[\text{mas}~\text{yr}^{-1}\right]$}
    & \multicolumn{2}{l}{$\left[\text{mas}~\text{yr}^{-1}\right]$}
    & & \\
    \hline
    2598522907060164480 & 22h40m34s
    & $-12^\circ33{}^\prime30^{\prime\prime}$ & $0.158$ & $\pm\!0.095$
    & $-0.17$ & $\pm\!0.09$ & $-0.6$ & $\pm\!0.1$ & 18.761
    & $\pm\!$0.007 \\
    2601287178076099712 & 22h31m39s
    & $-12^\circ50{}^\prime45^{\prime\prime}$ & $-0.004$
    & $\pm\!0.079$ & $0.05$ & $\pm\!0.08$ & $-0.8$ & $\pm\!0.1$
    & 18.014 & $\pm\!$0.004 \\
    2601407505880573696 & 22h36m13s
    & $-12^\circ53{}^\prime38^{\prime\prime}$ & $-0.046$
    & $\pm\!0.139$ & $-0.01$ & $\pm\!0.15$ & $-0.7$ & $\pm\!0.1$
    & 18.546 & $\pm\!$0.006 \\
    2601408089995818752 & 22h36m29s
    & $-12^\circ51{}^\prime13^{\prime\prime}$ & $0.136$ & $\pm\!0.132$
    & $0.04$ & $\pm\!0.16$ & $-0.6$ & $\pm\!0.2$ & 18.505
    & $\pm\!$0.006 \\
    2601835941752991872 & 22h34m33s
    & $-11^\circ47{}^\prime24^{\prime\prime}$ & $-0.064$
    & $\pm\!0.087$ & $0.01$ & $\pm\!0.10$ & $-0.5$ & $\pm\!0.1$
    & 18.047 & $\pm\!$0.005 \\
    \hline
  \end{tabular}
  \label{tab:targets}
\end{table*}

\subsection{Magellan Inamori Kyocera Echelle (MIKE) spectroscopy}%
\label{spectroscopy_subsection}

We observed the clustered 5 stars with Magellan/MIKE
\citep{Bernstein2003} on 2022 Jun 10 using the 0\farcs7 slit, 2x2
binning, and slow readout, resulting in approximate spectral
resolution $R\sim 22,000$ on the red arm. Exposure times ranged from
15-30 min per star in good (${<}0\farcs7$) seeing. Data were reduced
using CarPy \citep{Kelson2003} including sky subtraction and
wavelength calibration with adjacent ThAr lamps. We did not correct
the spectra for telluric absorption. The resulting median
signal-to-noise ratio was 16-26 per pixel at
$8560-8570~\mathrm{\mathring{A}}$ near the \ion{Calcium}{II} triplet
(CaT). 

\begin{figure*}
\includegraphics{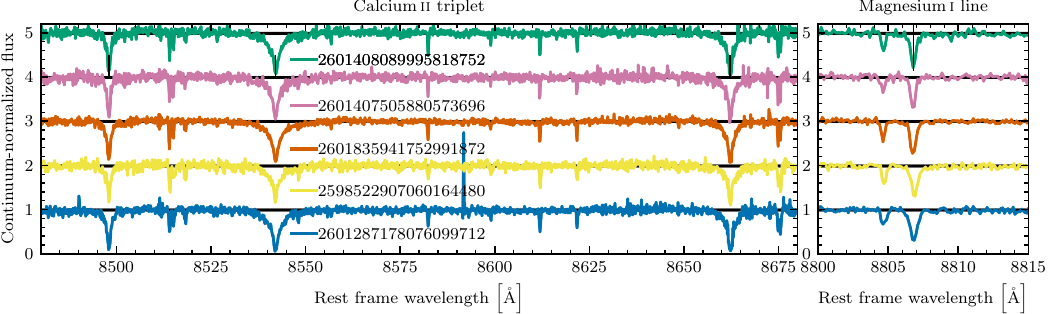}
\caption{Normalized MIKE spectra of the five target stars in the (rest)
         wavelength regions where the \ion{Calcium}{II} triplet and
         infrared \ion{Magnesium}{I} line are
         visible.}
\label{fig:spectra}
\end{figure*}

For the subsequent analysis of the regions around the CaT and
infrared \ion{Magnesium}{I} line at 8806.8~$\mathrm{\mathring{A}}$
(Mg8806) we continuum-normalized the spectra using the
\texttt{continuum} function in the \texttt{onedspec} package in the
Image Reduction and Analysis Facility
\citep{tody1986, tody1993, iraf}. For the parameters of the
\texttt{continuum} function we chose a cubic spline model and lower
2$\sigma$-clipping within ten iterations while also rejecting
neighboring pixels within a distance of one pixel to the rejected
pixels. The resulting spectra in the CaT and Mg8806 regions are shown in
\autoref{fig:spectra}. We note that CarPy does not correct pixels
affected by cosmic rays from single exposures. That is why there are
some parts of the spectra with spikes that result from pixels affected
by cosmic rays (see, e.\,g., fourth/green spectrum). However, we note
that these spurious pixels in the spectra do not affect the parts of the
spectrum that we analyze.
        
\section{Spectral analysis}
\label{sec_analysis}

\subsection{Full spectral fitting}\label{spectral_fit_subsection}

We performed initial spectral fits using a high resolution optical
version of The Payne \citep{Ting2019} as implemented in
\texttt{LESSPayne} \citep{Ji2025}. We perform a full spectrum fit from
$5000-7000\mathrm{\mathring{A}}$, simultaneously optimizing four
stellar labels (T$_\text{eff}$, $\log g$, [Fe/H], [$\alpha$/Fe]),
line-of-sight velocity, smoothing kernel, and per-order continuum
parameters. The fits suggested the five stars were all relatively
metal-rich giant stars ($\mbox{[Fe/H]} \sim -1.4$,
$\log g \lesssim 1$). We consider for all line-of-sight velocities
measured from MIKE spectra a systematic uncertainty floor of
1~km~s$^{-1}$ \citep{alex_p_ji_et_al2020}.

However, as the S/N is very low in this wavelength range and the
spectral model has not been properly calibrated at redder wavelengths
beyond $7000\,\mathrm{\mathring{A}}$, we perform a more focussed
spectral analysis in Section \ref{spectral_analysis_subsection}
focusing on just the higher S/N region around the CaT. Nonetheless, we
use the resulting [Fe/H] and line-of-sight velocity estimates as
initial guesses and reference values in the subsequent analysis. These
are given in \autoref{initial_spectral_fit_results}. We refrain from
reporting the [$\alpha$/Fe] estimates from the initial spectral fits
because we do not consider these estimates trustworthy.

\begin{table}
\caption{Results from the initial spectral fitting.}
\begin{tabular}{lll}
\hline
Source ID & $V _ \mathrm{h,initial}$
& $[\text{Fe}/\text{H}] _ \text{initial}$ \\
& $\left[\text{km}~\text{s}^{-1}\right]$ &  \\
\hline
2598522907060164480 & $-170.4$ & $-1.44$ \\
2601287178076099712 & $-214.9$ & $-1.46$ \\
2601407505880573696 & $-191.1$ & $-1.61$ \\
2601408089995818752 & $-208.3$ & $-1.33$ \\
2601835941752991872 & $-208.4$ & $-1.48$ \\
\hline
\end{tabular}
\label{initial_spectral_fit_results}
\end{table}

\subsection{Distinguishing between RGB stars and main-sequence dwarf
            stars}%
\label{spectral_analysis_subsection}

Due to the limited S/N, the initial full-spectrum stellar fitting is not
deemed a very reliable measure of surface gravity. In order to ensure
that we have not selected foreground main-sequence stars with similar
colours as the (expected) distant RGBs, we perform an additional
gravity-sensitive analysis on the much higher S/N region around 8500
$\mathrm{\mathring{A}}$. \citet{battaglia_starkenburg2012} advocate
that a combination of the equivalent width measurements of the second
and third infrared \ion{Calcium}{ii} triplet (CaT) lines in this region
when combined with the Mg8806 line in the same wavelength regime can
help to separate red giant from main-sequence dwarf stars, as these
lines show different behaviour with increasing surface gravity. We
apply this method to the five clustered stars by measuring the
equivalent widths of these three absorption lines. Figure
\ref{fig:spectra} shows these lines for each of our program stars.

In the inference of the equivalent widths of the CaT and Mg8806 we
adapt a pipeline that was originally developed by
\citet{longeard_et_al2022} and later modified by
\citet{viswanathan_et_al2024} and \citet{viswanathan_et_al2025}. For
our purposes, we fit the lines individually within a
15-$\mathrm{\mathring{A}}$ window each with Gaussian absorption line
profiles solving for their Doppler shift (and thus line-of-sight
velocity), normalized fluxes, depth, and standard deviations, and
taking into account the (local) continuum uncertainty.

The fitting is performed by a maximum-likelihood estimation technique
through a Markov chain Monte Carlo algorithm (MCMC, as implemented in
\texttt{emcee} and described in \citealt{foreman-mackey_et_al2013}). We
identify the set of parameters that optimizes the natural logarithm of
the Gaussian likelihood function of the observed spectrum and its
uncertainty, including the (local) continuum uncertainty given the free
parameters of the simulated spectrum, which is equivalent to minimizing
the chi-square between observed and simulated spectrum. There are some
intrinsic priors we assume here. For instance, the depth and width
parameters of the Gaussian absorption lines profile in the simulated
spectrum cannot be negative, and we require for the two CaT lines
that the second (middle) line is the strongest. Results from the
initial spectral fitting mentioned in Section
\ref{spectral_fit_subsection} are used as initial guesses for the
line-of-sight velocities. The step size of the algorithm, which is
defined for the purpose of exploring the parameter space to achieve the
optimal acceptance ratio, is determined based on the S/N of the
spectrum. The fitted equivalent widths are multiplied with a factor 1.1
for the CaT lines to incorporate the effect of the non-Gaussian wings
\citep[see for more details][]{%
  battaglia_et_al2008mnras, e_starkenburg_et_al2010},
and together with the measured line-of-sight velocities these are
provided in \autoref{spectroscopic_results} and well-constrained due to
sufficient S/N around the CaT.

\begin{figure}
\includegraphics{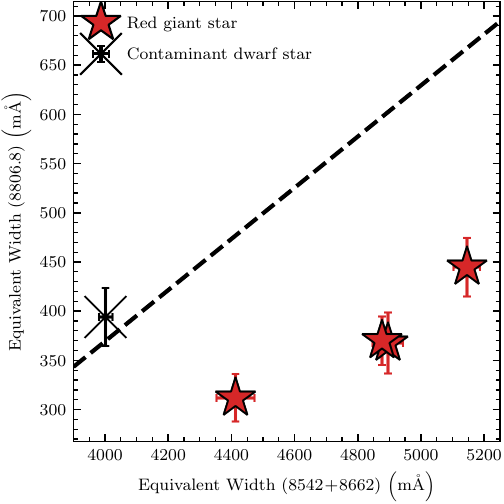}
\caption{Distinction of candidate giant and dwarf stars of the studied
         set of five stars with MIKE
         spectra using the method investigated in
         \citet{battaglia_starkenburg2012}. This method employs the
         summed equivalent widths of the two strongest lines of the
         \ion{Calcium}{ii} triplet at 8542~$\mathrm{\mathring{A}}$ and
         8662~$\mathrm{\mathring{A}}$ and the equivalent width of the
         Mg8806. We show part of the distinction line derived in
         \citet[][their equation 1]{battaglia_starkenburg2012} between
         giant and dwarf stars. \textit{Gaia} (E)DR3
         2598522907060164480 is highlighted because this star is
         discrepant compared to the other four stars and also close to
         the line.}
\label{equivalent_width_plot}
\end{figure}

Figure \ref{equivalent_width_plot} shows the resulting equivalent width
space as used in \citet{battaglia_starkenburg2012} together with the
diagonal distinction line that is drawn in this paper between the red
giant and main-sequence dwarf populations. It is striking that
\textit{Gaia} (E)DR3 2598522907060164480 is a candidate dwarf star
according to our method. Not only does it lie on the other side of the
distinctive line, it also clearly does not follow the trend of the other
four stars in equivalent width-space. While it can not be concluded with
certainty that this star is a main-sequence interloper -- some RGB stars
are found in the datasets of \citet{battaglia_starkenburg2012} close to,
or even just beyond, this line -- it certainly implies that it cannot be
determined that it is a bona fide outer halo red giant. We note that at
faint magnitudes, the Gaia parallax is not very
constraining either, as it is expected to have a large uncertainty even
if the star was a main-sequence star (see, e.g., \vb\
\citeyear{viswanathan_bystroem_et_al_arxiv2408_17250}, who see this
effect around $G > 17.3$). We furthermore note that \textit{Gaia}
(E)DR3 2598522907060164480 has the largest parallax
as well as ipd\_gof\_harmonic\_amplitude measured within the sample 
(see Table \ref{tab:targets} and
$ipd\_gof\_harmonic\_amplitude = 0.13$; see
\cite{gedr3, fabricius_et_al2021} for usage of
ipd\_gof\_harmonic\_amplitude) and additionally, that it has a very red
color in all available color bands beyond the predictions of isochrones
from the Bag of Stellar Tracks and Isochrones
\cite[BaSTI,][]{hidalgo_et_al2018, pietrinferni_et_al2021,
                pietrinferni_et_al2024}
given the metallicities and halo-like age of these stars. All these
factors increase the uncertainty of its classification. Considering
this significant uncertainty on the dwarf-giant classification of
\textit{Gaia} (E)DR3 2598522907060164480, we only consider the four
other stars as bona fide RGBs for the remainder of this work.
           
\begin{table*}
\caption{Results from the analysis of the \ion{Calcium}{II} triplet and
         Mg8806.}
\begin{tabular}{lllllllllll}
\hline
Source ID & $V _ \text{h}$ 
& \multicolumn{2}{l}{$\text{Equiv.\ width}_1$}
& \multicolumn{2}{l}{$\text{Equiv.\ width}_2$}
& \multicolumn{2}{l}{$\text{Equiv.\ width}_3$} & [Fe/H]
& \multicolumn{2}{l}{$\text{Equiv.\ width} _ \text{Mg}$}\\
& $\left[\text{km}~\text{s}^{-1}\right]$
& \multicolumn{2}{l}{$\left[\mathrm{\mathring{A}}\right]$}
& \multicolumn{2}{l}{$\left[\mathrm{\mathring{A}}\right]$}
& \multicolumn{2}{l}{$\left[\mathrm{\mathring{A}}\right]$} &
& \multicolumn{2}{l}{$\left[\mathrm{\mathring{A}}\right]$} \\
\hline
2598522907060164480 & $-171.0$ & 1.02 & $\pm\!0.01$ & 2.22
& $\pm\!0.02$ & 1.78 & $\pm\!0.02$ & nan & 0.41 & $\pm\! 0.12$ \\
2601287178076099712 & $-215.8$ & 1.35 & $\pm\!0.02$ & 2.86
& $\pm\!0.04$ & 2.29 & $\pm\!0.03$
& $-1.40^{+0.05(\text{stat}.)+0.10(\text{syst}.)}%
_{-0.10(\text{stat}.)-0.13(\text{syst}.)}$
& 0.27 & $\pm\! 0.09$ \\
2601407505880573696 & $-189.8$ & 1.13 & $\pm\!0.04$ & 2.40
& $\pm\!0.05$ & 2.01 & $\pm\!0.03$
& $-1.60^{+0.05(\text{stat}.)+0.11(\text{syst}.)}%
_{-0.12(\text{stat}.)-0.15(\text{syst}.)}$
& 0.18 & $\pm\! 0.10$ \\
2601408089995818752 & $-208.2$ & 1.25 & $\pm\!0.02$ & 2.76
& $\pm\!0.04$ & 2.14 & $\pm\!0.03$
& $-1.34^{+0.04(\text{stat}.)+0.09(\text{syst}.)}%
_{-0.12(\text{stat}.)-0.13(\text{syst}.)}$ & 0.36 & $\pm\! 0.11$ \\
2601835941752991872 & $-207.2$ & 1.28 & $\pm\!0.01$ & 2.73
& $\pm\!0.02$ & 2.14 & $\pm\!0.01$
& $-1.41^{+0.03(\text{stat}.)+0.10(\text{syst}.)}%
_{-0.11(\text{stat}.)-0.13(\text{syst}.)}$
& 0.23 & $\pm\! 0.07$ \\
\hline
\end{tabular}\\
$V _ \text{h}$ is inferred from the CaT. The uncertainites of the
equivalent widths also account for the (local) continuum uncertainty.
[Fe/H] has a component of statistical uncertainty that resulted from
propagating the uncertainties in observed magnitude, color, CaT
equivalent widths and heliocentric distance of the stars into the
calibration by \citet{e_starkenburg_et_al2010}. As this calibration
itself has a reported maximum error of 8\%, this additional component
is noted as the systematic uncertainty of [Fe/H].
\label{spectroscopic_results}
\end{table*}

\subsection{Isochrone fitting}\label{subsection_isochrone_fits}

\subsubsection{Distances for our four RGB stars}

For our four RGB stars, we derive their distances by combining
their (de-reddened) apparent magnitudes with their absolute
magnitude obtained from isochrones. In particular, we use here the
Pan-STARRS1 $(g\! -\! i)_0$-$g_0$ color-magnitude diagram
\cite[as also used for this purpose in e.\,g.,][]{%
  chambers_et_al_arxiv1612_05560, longeard_et_al2018,
  longeard_et_al2020}
and compare the placement of our four stars in this space with
predictions of the BaSTI isochrones
\citep{hidalgo_et_al2018, pietrinferni_et_al2021,
       pietrinferni_et_al2024}.
The choice for this set of isochrones is motivated in
\vb\ (\citeyear{viswanathan_bystroem_et_al_arxiv2408_17250}), where
it is shown that these isochrones produce the best agreement on the
RGB when compared to inverted-parallax distances measured by the
\textit{Gaia} space telescope \citep{gaia_mission, gdr3}. Observed
photometry was dereddened using the dust maps of
\citet[][renormalized by \citealt{schlafly_finkbeiner2011}]{sfd}.

Moreover, the BaSTI isochrones have the advantage that they are
provided for a range of different $[\alpha/\text{Fe}]$ values, from
depleted ($[\alpha/\text{Fe}]  = -0.2$;
\citealt{pietrinferni_et_al2024})
to enhanced ($[\alpha/\text{Fe}]  = 0.4$;
\citealt{pietrinferni_et_al2021}). For our purpose, we choose the
enhanced set, as it is expected that stars of this metallicity will be
$\alpha$-enhanced in the Sagittarius system \citep[see the left panel
figure 3 in][]{de_boer_et_al2014}. In the selected isochrones we
include stellar overshooting and diffusion, and set the free parameter
in the Reimers law of stellar mass loss to the recommended value of
$\eta  = 0.3$ \citep{cassisi_salaris2013}. For the metallicities of the
isochrones, we use the results from the full spectrum fitting presented
in Table \ref{initial_spectral_fit_results}. The final parameter for the
choice of isochrone for each star, is the stellar age. Although age has
a small effect at old ages, here we base our estimate on the age
analysis in the Sagittarius stellar stream in the lower right panels of
the figures 5 and 9 of \citet{de_boer_et_al2015}. We consider 13~Gyr as
the fiducial age, since the constrained star formation in the
Sagittarius stream by \citet{de_boer_et_al2015} peaks at this age for
the [Fe/H]-range of the four RGB stars. To obtain the absolute
magnitude expected for each of our four stars, and thus the distance
modulus, we determine the closest point on the isochrone RGB branch by
matching the $(g\! -\! i)_{\text{PS},0}$ colour.
 
We furthermore repeat this exercise for isochrones of 7~Gyr and 14~Gyr
as lower and upper limits, respectively
\citep[again based on the results by][]{de_boer_et_al2015}, and use the
results from these older and younger isochrones to determine (a major
element of) the uncertainty on the distance determination while noting
that the resulting distance for isochrones of old stellar ages around
13 and 14\,Gyr do not differ significantly at these distances.
Photometric color uncertainties can be neglected in their contribution
to the final distance uncertainties, because they are much smaller than
the range between the different plausible isochrones. We do, however,
add in quadrature a distance uncertainty floor of 12\%, matching the
average uncertainty reported in
\vb\ (\citeyear{viswanathan_bystroem_et_al_arxiv2408_17250}) when they
compared their distances -- obtained with a very similar method -- of a
sample of over 6 million RGB stars to inverted parallax results. We note
that this uncertainty floor also comfortably incorporates the expected
systematic uncertainties between different isochrone models implemented
here. While our reference ages from \citet{de_boer_et_al2015} used
Dartmouth isochrones \citep{aaron_dotter_et_al2008dartmouth}, we employ
BaSTI isochrones, but the two isochrone sets only differ by 
$\pm\! 2.2\%$ in the range of $(g\! -\! i)_{\text{PS},0}$ for our four
target stars for the same ages, [Fe/H], and $[\alpha/\text{Fe}]$.

Our final photometrically determined heliocentric distances of the four
RGB stars, and corresponding uncertainties, are given in the second
column of \autoref{isochrone_fits_results_table}. 

\begin{table}
\caption{Results from BaSTI isochrone fitting for our four RGB stars.}
\begin{tabular}{llllll}
\hline
Source ID & $D _ {\text{h},\text{iso}}$ & \multicolumn{2}{l}{$\log g$}
& \multicolumn{2}{l}{$T _ \text{eff}$} \\
& [kpc] &  &  & \multicolumn{2}{l}{[K]} \\
\hline
2601287178076099712 & $94^{+13}_{-11}$ & 0.3 & $\pm \! 0.1$ & 3934
& $\pm \! 31$ \\
2601407505880573696 & $106^{+18}_{-13}$ & 0.8 & $\pm \! 0.1$ & 4217
& $\pm \! 31$ \\
2601408089995818752 & $95^{+16}_{-11}$ & 0.7 & $\pm \! 0.1$ & 4100
& $\pm \! 31$ \\
2601835941752991872 & $84^{+14}_{-10}$ & 0.7 & $\pm \! 0.1$ & 4112
& $\pm \! 31$ \\
\hline
\end{tabular}
\label{isochrone_fits_results_table}
\end{table}

\subsubsection{Validation of the RGB distances}

\begin{figure}
\includegraphics{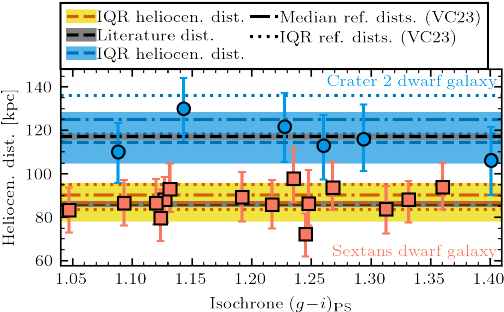}
\caption{Distance scale of the four RGB stars from the
         isochrone fits is being validated using stars from the
         Sextans and Crater 2 dwarf galaxies. The heliocentric,
         literature distances (horizontal, shaded, dark lines) are
         recovered within the interquartile range (horizontal,
         shaded lines where the dashed line marks the median). As a
         reference, we plot the median and interquartile range of the
         heliocentric distance estimates of the same set of stars in
         each dwarf galaxy by \citet{vedant_chandra_et_al2023b} that
         used the same sets of stars for their validation of the
         distance scale as dashed dotted lines.}
\label{fig:validation_isochrone_dists}
\end{figure}

To test the performance of our distance method, we use two datasets of
RGB stars in the Sextans and Crater 2 dwarf spheroidal galaxies
\citep[these data were also presented for distance validation in][]%
{vedant_chandra_et_al2023b}.
We apply the same methodology outlined above to these two datasets and
choose the BaSTi isochrone for each star best fitting the results
for the \texttt{MINESweeper} \citep{cargile_et_al2020} analysis for
[Fe/H] and [$\alpha$/Fe] performed in \citet{vedant_chandra_et_al2023b},
taking into account also the uncertainties on these values. The age
range for each Sextans and Crater 2  RGB star is based on constraints by
studies by \citet{bettinelli_et_al2018} and \citet{torrealba_et_al2016},
respectively. For Crater 2, we consequently adopt a fiducial isochrone
of 10~Gyr and adopt an uncertainty of 1~Gyr.
\citet{bettinelli_et_al2018} derive very old ages for all stars in
Sextans, but indicate that there might be a mild age-metallicity trend.
We therefore adopt an isochrone of 13.5~Gyr for Sextans RGB stars with
[Fe/H] $< -1.85$ and an age of 12.3~Gyr for more metal-rich stars in
Sextans (although we note that isochrones for such old ages tend to
follow each other very closely).

Figure \ref{fig:validation_isochrone_dists} shows the results
of these tests. Our estimates of the heliocentric distances for each
RGB star (shown as circles and squares) are in excellent agreement with
the literature values of the distances to both Sextans
\citep[$86 \!\pm\! 1$~kpc;][]{irwin_et_al1990, munoz_et_al2018} and
Crater 2 \cite[$117 \!\pm\! 1$~kpc;][]{torrealba_et_al2016}, both
indicated by black dashed lines and grey bands. The median distances
for each galaxy, indicated with coloured dashed lines track the
literature values almost perfectly and are in much better agreement
than the distance results for the same sets of stars by
\citet[][shown as dash-dotted and dotted lines]%
{vedant_chandra_et_al2023b}
based on MIST isochrones
\citep[MIST,][]{a_dotter2016, j_choi_et_al2016}. We see this as
confirmation that the $\alpha$-variable BaSTI set of isochrones is
indeed a better choice for these stars.

We additionally investigate whether there are is any trend with the
distance validity and the position of the star along the RGB, shown as
the x-axis of \autoref{fig:validation_isochrone_dists} by proxy of
$(g\! -\! i)_\text{PS}$. One could expect such differences for instance
if the slope of the model isochrones does not match that of the real
stars in the population. We see no significant trend in
\autoref{fig:validation_isochrone_dists} and this is confirmed by
Spearman rank test using a Monte Carlo approach to include also the
distance uncertainties. These tests indicated an average correlation
coefficient of 0.1 and -0.2 (for Sextans and Crater 2) between the
distance offset and $(g\! -\! i)_\text{PS}$, but with 16th and
84th quantiles of -0.2 and 0.4 and -0.7 and 0.4, respectively. We note
that instead a correlation is present if the distance uncertainty floor
of 12\% is not taken into account, thus we take these tests to show
that our distances are robust and moreover that our uncertainties can
be considered realistic.

\subsubsection{Other stellar parameters from isochrone fitting}

In addition to stellar distances, we also derive
stellar effective temperatures and surface gravities for our stars from
the isochrone fitting, using the same isochrones and procedure.
Spectroscopic estimates of $T _ \text{eff}$ and $\log g$ of RGB stars
with $[\text{Fe}/\text{H}] \lesssim -1.5$ are more
affected by 3D/NLTE effects than photometric estimates (e.g.,
\cite{mucciarelli_bonifacio2020}; see also \cite{amarsi_et_al2016}).
For both
$T _ \text{eff}$ and $\log g$ we use an uncertainty floor based on the
results from \citet{mucciarelli_bonifacio2020} who studied globular
clusters NGC\,5634, NGC\,6809, NGC\,6093, NGC\,1904, NGC\,6752,
NGC\,288, and NGC\,5904 and found median intrinsic dispersions of 31~K
and 0.1 for $T _ \text{eff}$ and $\log g$ between spectroscopic and
photometric estimates of RGB stellar parameters. They furthermore
report median offsets for $T _ \text{eff}$ and $\log g$ between
spectroscopic and photometric stellar parameters of -111\,K and
$-0.13$. We consider these uncertainty estimates as lower limits for
our study of field stars whose place on the isochrone is less
well-constrained than in a globular cluster system. However, we note
that many of the additional uncertainties will be systematic in nature
and we adopt the intrinsic dispersions above for our measurements that
are solely used for relative analysis (see Section
\ref{sub-subsection_disrupted_globular_cluster}).

The values for $T _ \text{eff}$ and $\log g$ as obtained through
isochrone fitting are given in
\autoref{isochrone_fits_results_table}.

\subsection{Metallicities}%
\label{iron_hydrogen_abundance_ratio_estimation}

As explained in \autoref{spectral_fit_subsection}, the reported
[Fe/H] from the full spectrum fitting are based on models that are not
calibrated well beyond 7000\,$\mathrm{\mathring{A}}$. We therefore
additionally carry out a more detailed analysis of the CaT to measure
stellar metallicities. In this work, we adopt the calibration of
\citet{e_starkenburg_et_al2010} for this purpose and combine the
Gaussian equivalent widths as presented in
\autoref{spectroscopic_results} (these values are including already a
factor 1.1 correction term for their non-Gaussian wings) together with
the absolute $M _ {I_\text{Cousins}}$ magnitude for the four RGB stars
to obtain their [Fe/H] values.%
\footnote{$I_\text{Cousins}$ was in turn estimated from the
          \textit{Gaia} (E)DR3 $G _ {BP,0}$, $G _ {RP,0}$ and $G _ 0$
          photometry using the last entry under Johnson-Cousins
          relationships in table C.2 in \citet{riello_et_al2021}.}
The resulting estimates of [Fe/H] given in
\autoref{spectroscopic_results} are in close agreement with the initial
estimates from the full spectral fitting presented in Table
\ref{initial_spectral_fit_results}. The mean offset between the two
methods is $\mu$ = 0.03, with dispersion $\sigma$ = 0.05. This is fully
consistent with the quoted measurement uncertainties. We stress that
both methods are completely independent, not only employing very
different techniques, but also using distinctly different wavelength
ranges of the spectra for these stars. We thus view this excellent
agreement as very encouraging, especially since our four RGB stars are
close to the tip of the RGB, a less-well checked and calibrated region
of the stellar parameter space.%
\footnote{We note that such an agreement is not met when we use the
          calibration of \citet{Carrera2013}. Instead, then we find a
          significant offset of $\mu$ = 0.41, with dispersion $\sigma$
          = 0.07. We suspect this might be due to the lack of stars
          close to the tip of the RGB in the sample that they use to
          derive the calibration. This would make the relation less
          constrained at the tip of the RGB, where our four RGB stars
          are located.}

To determine the measurement uncertainties associated with the
metallicities inferred, we employ a Monte Carlo procedure. For each
iteration, we randomly draw values of the equivalent widths from their
probability distribution functions. We then compute the corresponding
spectroscopic metallicity, taking into account also the individual
uncertainties for each star on its magnitude and distance. By repeating
this procedure 100 times, we construct a full probability distribution
function for the metallicity of each star from which we extract the
lower and upper bounds on the metallicity uncertainties as the 16th and
84th quantile values. These are the statistical uncertainties on the
metallicity as quoted in \autoref{spectroscopic_results}.

Using a similar a Monte Carlo approach we also include in our analysis
the uncertainties from the calibration relation itself by considering
the (conservative, upper limit of) 8\% systematic error quoted in
\citet{e_starkenburg_et_al2010}. We consider these resulting
uncertainties as more systematic in nature (if the calibration is
inaccurate, it is most likely affecting all stars in our sample in a
similar direction, so we argue this can not be treated as a component of
the statistical uncertainty) and we note it separately in
\autoref{spectroscopic_results}.

As can already be appreciated from \autoref{spectroscopic_results}, our
four stars show very similar metallicities. We investigate this further
-- and discuss whether this could mean we are looking at a Sagittarius
galaxy globular cluster remnant -- in Section
\ref{sub-subsection_disrupted_globular_cluster}.

\subsection{Embedding the cluster of four distant RGB stars in the
            broader Sagittarius stream}

In this section, we aim to collect all additional stellar data that is
currently available on the southern, distant Sagittarius stream to
compare it to the cluster of four RGB stars studied in this work.
In the following subsections, we describe our selection of RR Lyrae
stars and stars in the stellar distance value-added catalog SPDist as
part of the Dark Energy Spectroscopic Instrument
\cite[DESI,][]{desi_edr, desi_dr1} Milky Way Survey
\cite[MWS,][]{andrew_p_cooper_et_al2023, koposov_et_al2024,
              koposov_et_al2026}
DR 1. We note that we also searched for blue horizontal branch stars
belonging to this substructure in existing catalogs that are close to
the cluster of the four RGB stars, but we did not find any, probably
due to the limited distance extend of these catalogues (the list of
literature we checked is given in \autoref{appendix_section_bhb_stars}).

In all searches, we restrict ourselves to stars with heliocentric
distance beyond 45 kpc, a $B$ smaller than $\pm\! 20^\circ$
compared to the Sagittarius orbital plane, and a longitude range of
$-122^\circ < \Lambda < 15^\circ$ on the orbital plane of Sagittarius.

\subsubsection{RR Lyrae stars}\label{sub-subsection_rr_lyrae}

We collected a set of (candidate) RR Lyrae stars of Bailey types ab and
c \citep{solon_irving_bailey1902} from two sources:
\begin{enumerate}
  \item the Pan-STARRS1 sample by \citet{sesar_et_al2017a} of class ab
  stars that have a final classification score higher than 80\%
  \citep[following other studies, such as,]%
  [for the classification score]{fardal_et_al2019}.
  \item dataset of ab- and c-type RR Lyrae stars from the \textit{Gaia}
  Data Release 3 catalog \citep{gaia_mission, clementini_et_al2023,
                                muraveva_et_al2025}.
\end{enumerate}
For the stars that are present in both catalogs we decided to use the
distance moduli in the table by \citet{sesar_et_al2017a}. Other sources
of RR Lyrae stars were searched
\citep{%
  kuan-wei_huang_koposov2022, yuting_feng_et_al2024, medina_et_al2024},
but did not have overlap with the area on sky of the four RGB stars.

We find a total of 22004 stars following this selection with reported
(heliocentric) distances beyond 45 kpc that we will investigate further
in Section \ref{distant_sgr_stream_subsection}.

\subsubsection{Red gaints from DESI value-added catalog SPDist}%
\label{spdist_sub-subsection}

We additionally searched a complementary set of distant RGB stars in
the value-added stellar distance catalog of the DESI DR1 MWS. We note
that DESI has not targeted the area of the sky in which we found the
four clumped RGB stars in their DR1, and thus there is no overlap of
targets. However, DESI includes pointings as close as 0.41$^\circ$ from
the location of our stars, making it perfectly suitable to search for
additional candidates of a broader Sagittarius spur feature. Thanks to
its spectroscopic nature, DESI provides full 6D phase space data these
stars. Besides our sky location cuts already mentioned, we use the
following constraints for candidate stars:

\begin{itemize}
  \item are in \textit{Gaia} DR3 
  \item have $\texttt{PRIMARY}  = \texttt{True}$
  \cite[recommended in][to exclude duplicate observations]%
  {koposov_et_al2024}.
  \item have $\texttt{RVS\_WARN}  = 0$
  \cite[recommended in]%
  [to select objects that have good fits in \texttt{RVSpecFit}]%
  {koposov_et_al2024}.
  \item are classified as stars based on the \texttt{Redrock}
  classification in DESI
  ($\texttt{RR\_SPECTYPE}  = \texttt{Star}$;
  \textcolor{blue}{Bailey et al., in preparation}).
  \item have $\texttt{SN\_R} > 10$ in the \texttt{RVSpecFit}
  pipeline of DESI MWS
  \item have $\texttt{VSINI} < 30$
  \cite[used also in][]{koposov_et_al2026}.
  \item have \texttt{FLAG\_GOOD} (recommendeded use of SPDist).
  \item have $\texttt{SNR\_MED} > 5$ and $\texttt{CHISQ\_TOT} < 5$ in
  the pipeline of DESI for
  the stellar parameters \cite[recommended in][]{koposov_et_al2024}.
  \item have $3500 < \texttt{TEFF}~\text{K}^{-1} < 5500$ and
  $\log g \leq 3.5$ in the 
  pipeline of DESI for stellar parameters to select red giant stars
  \cite[reduced lower limit of effective temperature recommended in]%
  [to include the brightest portion of the RGB]{koposov_et_al2024}.
  \item have $3300 < \texttt{TEFF}~\text{K}^{-1} < 5500$ and
  $\log g \leq 3.5$ in the RVSpecFit pipeline of DESI to select red
  giant stars.
  \item have $\texttt{BESTGRID} \neq \texttt{s\_rdesi1}$ in the
  pipeline of DESI for stellar parameters
  \cite[recommended in]%
  [to avoid stars that were fitted with the PHOENIX grid ]%
  {koposov_et_al2024}.
  \item have $\texttt{FEH} > -3.9$ in the \texttt{RVSpecFit}
  pipeline of DESI
  \cite[recommended in]%
  [to avoid stars at the edge of the grid]{koposov_et_al2024}.
\end{itemize}
We also follow the recommended use of SPDist in terms of distance
computations, meaning that we take the median of the absolute magnitude
$M _ G$ distribution in \textit{G} and get the uncertainties for $M_G$
from the $1\sigma$ quantiles (16th and 84th) plus in quadrature the
intrinsic precision of the method used to estimate the absolute
magnitudes of stars in SPDist (0.167, 8\% relative distance
precision).

After all these quality cuts, we add another 165 stars from this 
dataset to our sample of distant tracers with reported
(heliocentric) distances beyond 45 kpc in the relevant area of the
sky where we expect the Sagittarius Southern spur.

\section{Results}\label{sec_results}

\subsection{Significance of the four clustered RGB stars}%
\label{results_significance}

The metallicities we derive for the four RGB stars are very close
to the peak metallicity of the Milky Way halo
(e.g., \citealt{jelte_de_jong_et_al2010, sesar_et_al2011,
                allende_prieto_et_al2014, deason_et_al2018,
                lancaster_et_al2019, conroy_et_al2019,
                sarah_a_bird_et_al2021, gaochao_liu_et_al2022};
 \vb\ \citeyear{viswanathan_bystroem_et_al_arxiv2408_17250}).
This raises the question if the four stars represent a significant
clustered overdensity, or if this could simply be a chance clumping
in the stellar halo in sky coordinates and kinematics. To
investigate this and quantify the probability of
these four stars to be a chance cluster belonging to the overall
Milky Way halo, we utilize the Gaia Universe Model Snapshot
\cite[GUMS,][]{a_c_robin_et_al2012} of the Milky Way (version 10).
We consider only stars/data points in the GUMS with:
\begin{itemize}
  \item population parameter $\text{Pop} = 3$ (halo),
  \item $-100^\circ%
  \leq \text{Right\ ascension\ (wrapped\ at\ }180^\circ)%
  \leq 48^\circ$,
  \item $G \leq 20.5$ (stars that are in principle observable by
  \textit{Gaia}),
  \item $1.3 < \left(G_{BP}\!-\!G_{RP}\right)_{Gaia} < 3.0$
  \cite[selection of red stars as done in][]{%
    vedant_chandra_et_al2023b},
  \item $\log g \leq 3.5$;
  \cite[giant stars following][]{vedant_chandra_et_al2023a}
  \item a distance compatible with our four stars of
  $74 < D_\text{h}~\text{kpc}^{-1}\leq124$
  \item $M_G\leq -2.09$ (tip of the RGB as the four RGB stars in
  our study)
  \item and latitude-like angle $B$ in the Sagittarius coordinate
  system smaller than $\pm\! 20^\circ$.
\end{itemize}

There are $1328 \!\pm\! 37$ stars (assuming simple Poisson
uncertainties) in the selection, but they span a much larger area on
sky than our four stars (the selection is deliberately chosen over a
larger area as we are interested in average properties of the
population and would like to move away from small number
statistics). This number would result in a density of stars of
$0.180\!\pm\! 0.005~\text{kpc}^{-1}~\text{deg}^{-1}$ across
declination, which is relatively close to the estimated density
across declination for our four RGB stars of
$0.151~\text{kpc}^{-1}~\text{deg}^{-1}$. While we note that folding
in all uncertainties on distances and magnitudes could increase
this by a factor of three, or decrease it to almost zero, our
general conclusion is that we roughly expect a similar background
density of halo stars at the tip of the RGB in this part of the
halo as we have found observationally with only these four RGB
stars.

We stress, however, that this analysis has not yet folded in any of
the kinematical information. These four stars are not only clumped
on the sky and in distance, they also provide a clump in proper
motion space and are very tightly clumped in line-of-sight
velocities (as well as in metallicity). On the other hand, the tip
of the RGB stars as selected above in GUMS have a very broad line-%
of-sight velocity dispersion of $120~\text{km}~\text{s}^{-1}$
(their
$\text{average\ line-of-sight\ velocity}%
= -43~\text{km}~\text{s}^{-1}$).
When we estimate the Epanechnikov kernel \citep{epanechnikov1969}
of the distribution of line-of-sight velocities, we find that the
combined probability of finding four stars at the tip of the RGB
with the same line-of-sight velocities as the set of four RGB stars
can be maximally on the order of $3 \!\times\! 10^{-10}$\% when we
sample 5000 sets of the four line-of-sight velocities of the four
RGB stars given the 1\,km\,s$^{-1}$ uncertainty of each line-of-%
sight velocity in a Monte Carlo approach.

In summary, we conclude that it is highly unlikely that these
stars present a chance clumping of field halo RGB stars.

\subsection{Connection of the clustered stars to Sagittarius stream
            debris}%
\label{results_simulationcomparison}

\begin{figure}
  \includegraphics{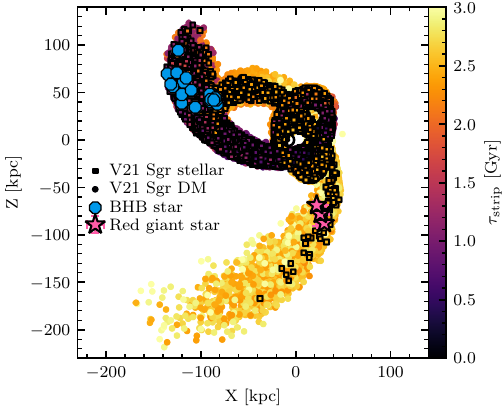}
  \caption{Galactocentric, Cartesian X-Z plot of the Sagittarius
           stream. We contextualized the group of four RGB stars
           within the broader framework of the Sagittarius stream and
           its predicted distant spur features in the Galactocentric
           north and south directions, as delineated in the simulation
           by \citet{vasiliev_et_al2021}. This simulation also tracks
           the escape time of each particle from its host galaxy, as
           depicted in the color scheme. We also include the set of
           blue horizontal branch stars (octagons) that were identified
           in the distant, northern spur in
           \citet{manuel_bayer_et_al2025}.}
  \label{xz_plot}
\end{figure}

\begin{figure*}
\includegraphics{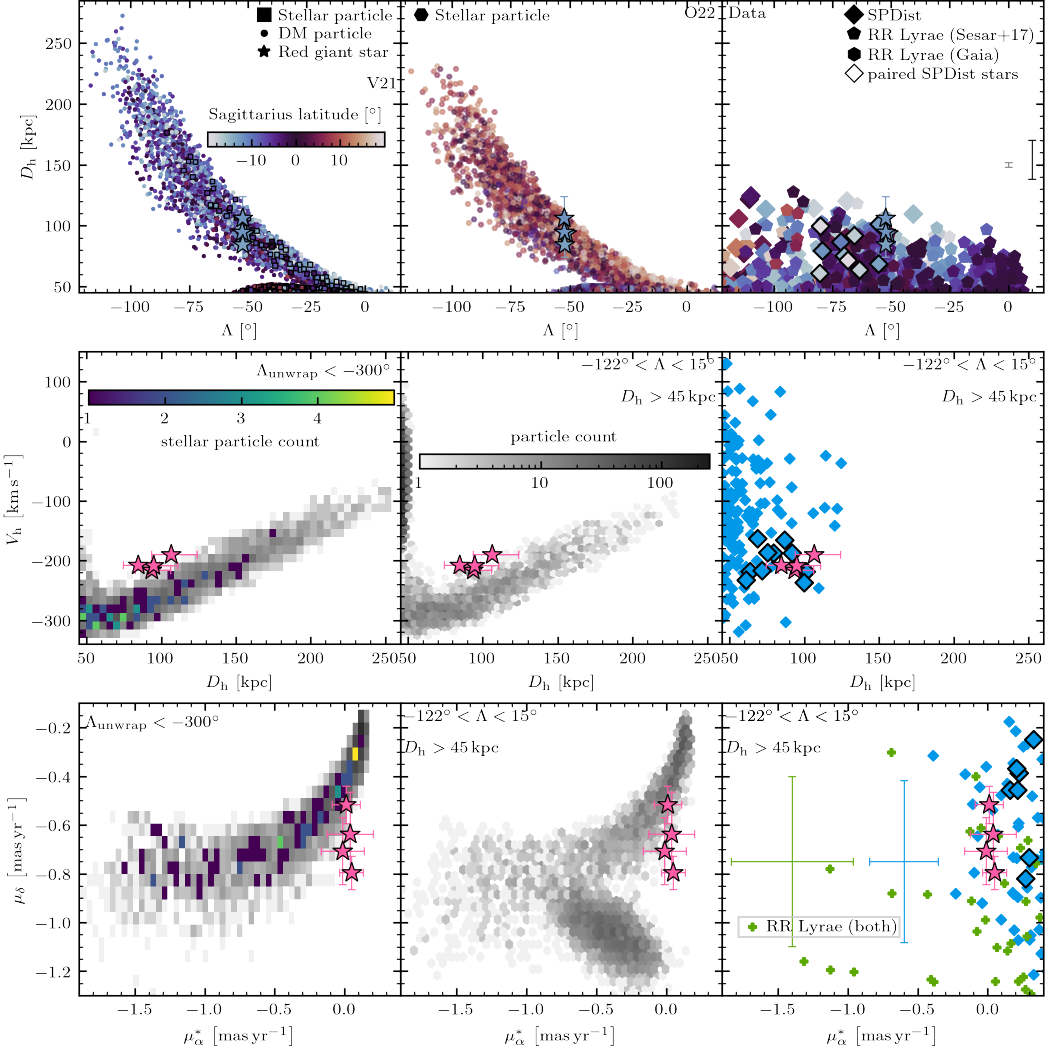}
\caption{Positional and motion data of the Sagittarius stream distant,
southern spur feature. The first and second column of panels show the
final snapshots of the simulations by \citet{vasiliev_et_al2021} and
\citet{oria_et_al2022}, respectively. The right column of panels shows
our data compilation.  The data compilation of stars includes RR Lyrae
stars from two catalogs: \citet{sesar_et_al2017a} (pentagon markers) or
\citet{muraveva_et_al2025} (hexagons) where stars present in both
catalogs are shown with pluses. Additional stars from the SPDist
catalogue of DESI DR1 \citep{desi_dr1} are shown as diamonds, where
stars close in position as well as distance, proper motion and
line-of-sight velocity (see text in Section
\ref{distant_sgr_stream_subsection} for details) are highlighted
with black edges. Colours in the top panels represent the latitude-like
angle, $B$, in the Sagittarius coordinate system, all panels only
includes data of stars with $B <$ $\pm\! 20^{\circ}$. Mean
uncertainties of heliocentric distance of RR Lyrae (bar markers in the
top-right panel) and SPDist stars (black bar marker in the top right
panel), proper motions of RR Lyrae (green marker) and SPDist stars
(blue marker) in the bottom-right panel are shown as well. The labels
in the lower two rows of panels provide the criteria we used in the
data or simulation to select the distant, southern spur feature.}
\label{sgrfar_data_figure}
\end{figure*}

As mentioned in the Introduction and Section
\ref{target_selection_subsection}, it was quickly realised that the four
RGB stars matched predictions from \citet{vasiliev_et_al2021} for the
Southern Sagittarius spur. Figure \ref{xz_plot} and the left and middle
panels in Fig. \ref{sgrfar_data_figure} illustrate this connection,
taking into account also the derived line-of-sight velocities from the
MIKE spectra, and show the four red gaints (star symbols) overplotted on
the final snapshot of a modeled Sagittarius stream. Figure
\ref{xz_plot} and the left panels of Fig. \ref{sgrfar_data_figure} show
the disruption of the Sagittarius dwarf galaxy in the time-dependent
gravitational potential of both the Milky Way and Large Magellanic
Cloud by \citet{vasiliev_et_al2021}, whereas the middle panels in Fig.
\ref{sgrfar_data_figure} show the simulation by \citet{oria_et_al2022}
(middle column). Only a limited distance and $\Lambda$ range are shown
to focus on the distant, southern spur of the Sagittarius stream, as
indicated in the relevant panels.%
\footnote{These are $-122^\circ < \Lambda < 15^\circ$ and
          $D_\text{h} > 45$~kpc in case of the data and the simulation
          by \citet{oria_et_al2022}. Since the authors of 
          \citet{vasiliev_et_al2021} provided an additional unwrapped
          longitude-like angle $\Lambda _ \text{unwrap}$, it is easier
          to select this feature in the simulation by
          \citet{vasiliev_et_al2021} through
          $\Lambda _ \text{unwrap} < -300^\circ$.}

It is important to note that the two simulations presented here are
not at all independent; \citet{oria_et_al2022} add a disk component in
the progenitor of the Sagittarius stream in their simulation, but use
otherwise similar initial conditions for their simulation setup. The
fact that, nevertheless, the predictions by the simulation by
\citet{vasiliev_et_al2021} fit better the observed latitude-like angles
$B$ in the coordinate system of the Sagittarius stream of the four
RGB stars, as shown in the colour coding of the top panels of Fig.
\ref{sgrfar_data_figure}, shows that this part of the stream is thus
very sensitive to the progenitor properties - and hence also observing
stars in these regions should provide potential discriminative power to
constrain these progenitor
properties.

In both simulations, the four RGB stars are overlapping broadly with
the simulation predictions for all parameters, although they are not in
the center of the parameter space occupied by the (stellar) simulation
particles. We will discuss this in more detail in Section
\ref{sec_traceback} where we also trace the simulation back to its
initial conditions, but we note here that especially the heliocentric
line-of-sight velocity and also proper motions across declinations are
tracing the outskirts of the trends of the dark matter and/or stellar
particles (for the results in \citealt{oria_et_al2022}, only the
stellar particles are available). The spread of the heliocentric,
line-of-sight velocities of the stellar particles in the spur
is larger in the simulation by \citet{oria_et_al2022}, which could be
explained by the addition of a disk component in the progenitor that
increases the range of angular momenta of the stellar particles.

\begin{figure}
  \includegraphics{%
    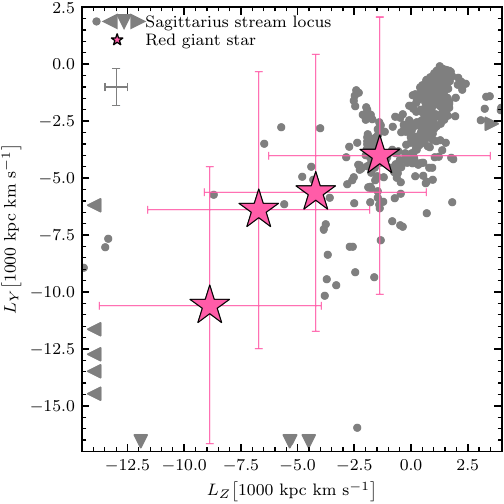}
  \caption{2D projection of the Galactocentric specific angular
           momentum space $\left(L _ Z\text{ versus\ } L _ Y\right)$
           (per mass) of the Sagittarius stream locus. We obtained the
           Sagittarius stream locus from identified candidate member
           stars in the BOSS-MINESweeper catalog
           \citep{vedant_chandra_et_al2026} of more nearby Sagittarius
           stream arms. Some of these stars that are outside of the axes
           limits are shown with arrow symbols. The average uncertainty
           of the Sagittarius stream locus is represented by the top
           left marker.}
  \label{specific_angular_momentum_plot}
\end{figure}

Besides a comparison with simulation predictions, angular momentum
space has also been used in the literature to identify Sagittarius
stream members
(e.g., \citealt{naidu_et_al2020, Johnson2020, Penarrubia2021,
                thomas_and_battaglia2022}; \vb\
\citeyear{viswanathan_bystroem_et_al_arxiv2408_17250};
\citealt{vedant_chandra_et_al2026}).
We show $\left(L _ Z\text{ versus\ } L _ Y\right)$ for our four RGB
stars in Fig. \ref{specific_angular_momentum_plot} overplotted on stars
identified as part of (more local arms of) the Sagittarius stream
selected in the BOSS-MINESweeper catalog by
\citet{vedant_chandra_et_al2026}. The four RGB stars in the cluster
follow broadly the Sagittarius stream locus, albeit with large
uncertainties due to their uncertain proper motion measurements. This
agreement strengthens their association to the Sagittarius stream.

Even though the morphology of the feature traced by our clump of
four RGB stars is not clear yet, we will for the remainder of this
work call it the 'southern spur', in line with the predictions from
the simulations and their (broad) match to the observed properties
of these stars. 

\subsection{A view on the Southern Sagittarius spur}
\label{distant_sgr_stream_subsection}
  
The right column of panels of \autoref{sgrfar_data_figure} shows
the results of our data collection in various parameter combinations
of sky position (all panels include the cut on $\pm\! 20^\circ$
compared to the Sagittarius orbital plane), distance, line-of-sight
velocity and \textit{Gaia} DR 3 proper motions. While none of the
additional stars is as close in the multi-dimensional space of
position and motion to our four RGB stars as they are to each
other, we are interested to see if in the other stars we can
distinguish a broader spur feature at this part of the sky as would
be suggested by the simulation predictions. 
 
When combining proximity in distance, sky position, and line-of-%
sight velocity to our clump by restricting to stars with
$D _ \text{h} > 60$~kpc,
$-250 < V_\text{h}~\text{km}^{-1}~\text{s} < -160$,
$\text{R.A.\ wrapped\ at}\ 180^\circ < 15^\circ$, we specifically
identify twelve stars in the set of SPDist stars to be candidate
members of a larger and approximately co-moving structure. Of these
twelve stars, ten also have similar proper motions (within
2$\sigma$) to our four RGB stars. This is a remarkably high
percentage, as for the overall SPDist sample at
$D _ \text{h} > 60$~kpc, this fraction is only 37\%. It
strengthens the hypothesis that there is a coherent motion at this
part of the sky at these distances as one would expect for a
Sagittarius spur feature.

We highlight these ten stars with black outlines in
\autoref{sgrfar_data_figure} and present their properties in 
\autoref{spdist_table} in \autoref{spdist_appendix_section}. Their
calibrated [Fe/H] from the DESI pipeline for Stellar Parameters
show a significant spread. While the metallicities for 7 of the
stars agree with the clump of four RGB stars within their
uncertainties, the 3 other stars do not. They do however all agree
(with the exception of 2601408089995818752) with the reported range
of [Fe/H] for RR Lyrae stars in the northern far arm of the
Sagittarius stream \citep{muraveva_et_al_arxiv2505_20165}. Because
the northern far arm of the Sagittarius stream would be composed of
stars that were lost from the progenitor at a similar time as in the
southern spur according to the simulation by
\citet{vasiliev_et_al2021}, this would be the most fair comparison
sample available, as the Sagittarius dwarf galaxy (and its streams)
possess a metallicity gradient (e.\,g.,
\citealt{s_l_j_gibbons_et_al2017},
\citealt{limberg_et_al2023}, \citealt{emily_c_cunningham_et_al2024},
\citealt{muraveva_et_al_arxiv2505_20165}, and Garc{\'i}a~Jim{\'e}nez
et al. in prep.).

We note that in future simulations with more realistically (and fully)
modeled stellar populations in the progenitor Sagittarius galaxy it
would be very interesting to investigate the expected number of RGB
stars in this southern spur and compare them quantitatively to
observations. Currently, we refrain from such an analysis as it would
require too many a-posteriori assumptions on the stellar counterparts of
the simulated dark matter particles to result in any meaningful
predictions (see also \autoref{sec_traceback}).

\section{Discussion}\label{sec_discussion}

\subsection{Stars from a disrupted globular cluster?}%
\label{sub-subsection_disrupted_globular_cluster}

Given that four RGB stars cluster closely in position and
kinematics, combined with their association to the Sagittarius
stream, we investigate the possibility that this tight group could be
part of a disrupted globular cluster - either disrupted by its pre-%
processing within the Sagittarius dwarf galaxy before infall, or
disrupted during the infall process. As such, it might be still visible
as a tight(er) clump inside a broader spur feature.

When we estimate the intrinsic line-of-sight velocity
$\sigma_{v_\text{los}}$ and metallicity dispersion
$\sigma_\text{[Fe/H]}$, we consider the uncertainty in the individual
stars during the fitting process of the observed line-of-sight
velocities and metallicities with a Gaussian model that incorporates
both measurement uncertainties and intrinsic dispersion. We utilize the
heliocentric line-of-sight velocities and metallicities from
\autoref{spectroscopic_results}. For the analysis of the metallicites we
do not use the systematic part of the metallicity uncertainty, as these
are expected to create merely offsets in the metallicity value rather
than impact the metallicity dispersion. We thus only use the asymmetric
measurement uncertainty associated to each star. To properly account for
these asymmetric uncertainties, we define an effective measurement
uncertainty for each star that depends on whether the model metallicity,
$\mu _ \text{[Fe/H]}$, lies above or below the measured value: 
\begin{equation}
  \sigma_i^\text{eff} =
  \begin{cases}
    \sigma_{i,-} & \text{if } \mu _ \text{[Fe/H]} < [\text{Fe/H}]_i.\\
    \sigma_{i,+} & \text{if } \mu _ \text{[Fe/H]}\geq [\text{Fe/H}]_i.
  \end{cases}
\end{equation}
The Gaussian model ensures that the total variance for each star is
accurately calculated through $\sigma_i^2 = \Delta_i^2 + \sigma^2$,
where
$\Delta_i\in \left\{\Delta V_{\text{h},i},\ \sigma_i^\text{eff}\right\}$
and
$\sigma\in\left\{\sigma_{v_\text{los}},\ \sigma_\text{[Fe/H]}\right\}$.
The ln-likelihood function is therefore
\begin{equation}
\ln \mathcal{L} = -\frac{1}{2} \sum_{i=1}^{N} \left[%
\frac{\left(x_i -\langle x\rangle\right)^2}{\sigma_i^2}%
+ \ln (2\pi\sigma_i^2)\right],
\end{equation}
where
$x_i\in\left\{V_{\text{h},i},\ [\text{Fe}/\text{H}]_i\right\}$
and
$\langle x\rangle%
\in\left\{%
\langle V_\text{h}\rangle,\ \mu_{[\text{Fe}/\text{H}]}\right\}$
is the average line-of-sight velocity or metallicity.

We sample the posterior distributions of $\langle x\rangle$ and
$\sigma$ using the \texttt{emcee} MCMC algorithm and uniform prior
probability distributions of $\langle V_\text{h}\rangle$
and $\sigma_{v_\text{los}}$ in the intervals (-216, -189)\,km\,s$^{-1}$
(range of the line-of-sight velocities of the four RGB stars)
and (0, 20)\,km\,s$^{-1}$, and $\mu_\text{[Fe/H]}$ and
$\sigma_\text{[Fe/H]}$ in the intervals (-5, 1) and (0, 2),
respectively, with five walkers exploring a two-dimensional parameter
space and running for 50000 steps. The first 100 steps of each chain
are discarded as burn-in, and the chains are thinned by a factor of 10
to reduce autocorrelation. Walkers are initialized near the average and
standard deviation of the observed line-of-sight velocities and
metallicities. The resulting posterior distributions naturally allows
$\sigma$ to approach zero, providing an upper limit on the intrinsic
dispersion when the measured metallicities are consistent with being
unresolved. The 50th percentile of the posterior is adopted as the
best-fit value, and the 16th and 84th percentiles are used to define
the 1$\sigma$ credible interval.

The posterior distributions of the intrinsic line-of-sight dispersion
for the four RGB stars in the distant, southern spur of the Sagittarius
stream give the best-fit values indicated by the 50th percentiles:
$\sigma_{v_\text{los}} = 12^{+5}_{-4}\,\text{km}\,\text{s}^{-1}$.
While at face value the four stars would follow the expected
velocity structure trend from the simulations by
\citet{vasiliev_et_al2021} and
\citet{oria_et_al2022} with an offset though in heliocentric
line-of-sight velocity as described in
\autoref{results_simulationcomparison}, we also considered the
line-of-sight velocity dispersion after excluding the most outlying
member on the sky and in distance (leaving open the possibility that
there might be a velocity trend in a globular cluster stream that is not
captured in the simulations). In this case the line-of-sight
velocity dispersion gets even smaller to
$\sigma_{v_\text{los}} = 7^{+6}_{-3}\,\text{km}\,\text{s}^{-1}$. This
would certainly be comparable to velocity dispersions observed for
globular clusters
\citep{william_e_harris1996, mclaughlin_and_van_der_marel2005,
       baumgardt_and_hilker2018, baumgardt_et_al2019_482_5138b}
and globular cluster streams
\citep{ting_s_li_et_al2021, ting_s_li_et_al2022,
       zhen_yuan_et_al2022, awad_et_al2024, viswanathan_et_al2025}
although we stress that this is a velocity dispersion from 3 (or 4)
stars and therefore caution should be taken to over-interpret these
results.

\begin{figure}
  \includegraphics{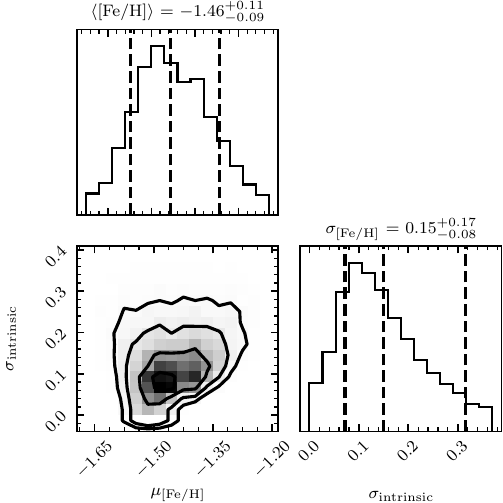}
  \caption{Constraints on the [Fe/H] average and dispersion of the
           four RGB stars in the cluster. We show the median and
           $1\sigma$ quantiles (16\% and 84\%) as dashed lines in
           the projections of the sampled 2D posterior probability
           distribution where the titles in the projections give the
           values. We used the function \texttt{corner} in the
           \texttt{corner} package to generate this plot
           \citep{foreman-mackey2016}.}
  \label{iron_hydrogen_abundance_ratio_dispersion}
\end{figure}

The posterior distributions of the mean metallicity and intrinsic
metallicity dispersion for the four RGB stars in the distant,
southern spur of the Sagittarius stream are shown in
\autoref{iron_hydrogen_abundance_ratio_dispersion}, with the best-%
fit values indicated by the 50th percentiles:
$\langle \text{[Fe/H]} \rangle = -1.46^{+0.11}_{-0.09}$ and
$\sigma_\text{[Fe/H]} = 0.15^{+0.17}_{-0.08}$ dex. The posterior
distribution for $\sigma_\text{intrinsic}$ extends to zero,
reflecting that the intrinsic dispersion is only weakly constrained
by our small sample and providing an upper limit on the metallicity
spread. The mean metallicity of this distant, southern Sagittarius
stream population is lower than that of the main body of the
Sagittarius dwarf spheroidal galaxy \cite[e.g.,][]{cole2001} and
also the values $-1.18\lesssim [\text{Fe}/\text{H}]\lesssim -1.03$
we expect from the constrained Sagittarius stream [Fe/H] gradient
across the Sagittarius latitude-like angle $B$ by
\citet{emily_c_cunningham_et_al2024}%
\footnote{\citet{emily_c_cunningham_et_al2024} uses the [M/H] notation,
          which are actually in their case on the same scale as [Fe/H]
          because they used the \citet{andrae_et_al2023} XP [M/H]
          metallicity estimates that were trained on Apache Point 
          Observatory Galactic Evolution Experiment (APOGEE) DR17
          \citep{majewski_et_al2017, abdurrouf_et_al2022_sdss_dr17}
          [M/H]. APOGEE [M/H] are on scale with [Fe/H]},
suggesting that these stars were stripped early in the disruption
process and may originate from the progenitor halo of Sagittarius
rather than its central, more metal-rich regions. We note, however:
first, the [Fe/H]($B$) linear model of
\citet{emily_c_cunningham_et_al2024} has a spread of
$0.286\!\pm\! 0.002$. Second, the \citet{andrae_et_al2023} [Fe/H]
estimates used by \citet{emily_c_cunningham_et_al2024} are
overestimated at $[\text{Fe}/\text{H}] < -1.0$ in comparison to
measurements by the Apache Point Observatory Galactic Evolution
Experiment DR17
\citep{majewski_et_al2017, abdurrouf_et_al2022_sdss_dr17}. The
correction of this bias will shift the predicted [Fe/H] based on the
gradient closer to our $\langle \text{[Fe/H]}\rangle$ and make it
more coherent.

The limit on the intrinsic spread in metallicity is consistent with
zero within $2\sigma$. This low metallicity dispersion is comparable to
that measured for several Milky Way globular cluster streams, such as
300S, Willka Yaku, Jet, Phoenix, and GD-1 \citep{ting_s_li_et_al2022},
consistent with a possible globular cluster origin. Taken together, the
mean metallicity and dispersion measurements suggest that the distant
Sagittarius stream group might represent material stripped from a
globular cluster of the Sagittarius progenitor halo, but we stress that
these results are very far from being convincing or conclusive and
further data would be necessary.

\begin{figure*}
  \includegraphics{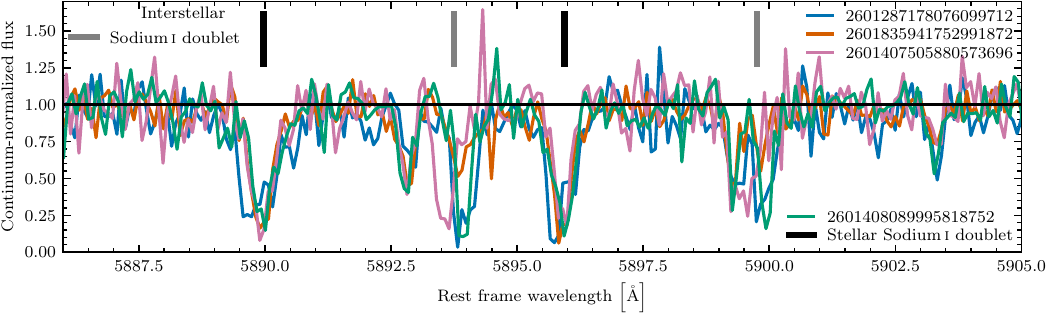}
  \caption{Normalized MIKE spectra of the four RGB stars in the
           wavelength region where the \ion{Sodium}{I} doublet 
           is visible. We also indicate the location of the ISM
           counterpart of these lines, that are offset from the
           stellar absorption lines.}
  \label{fraunhofer_sodium_doublet_figure}
\end{figure*}

To further investigate the speculative globular cluster origin of
the four RGB stars in the distant Sagittarius stream spur, we
examined sodium abundance variations, which are commonly observed
in Milky Way globular clusters as part of the Na-O anticorrelation
\citep[e.g.,][]{2009carretta, 2012gratton, nate_bastian_and_lardo2018}.
Visual inspection of the spectra in the Na~D region (see
\autoref{fraunhofer_sodium_doublet_figure}) shows no obvious
abundance variations. We quantified this by fitting the sodium
lines using \texttt{Korg} \citep{2023korg} with stellar parameters
taken from \autoref{isochrone_fits_results_table} and metallicities
from \autoref{spectroscopic_results}. We assumed a S/N estimated at
the local continuum near 5890~$\mathrm{\mathring{A}}$ and used the
MIKE spectrograph resolution ($R\sim22{,}000$) in our fits.
The four stars in the distant, southern feature of the Sagittarius
stream exhibit consistent sodium abundances, with a mean value of
$\langle A(\text{Na})\rangle = -1.68 \pm 0.03 \, \text{dex}$ in LTE
and $-1.56 \pm 0.03$\,dex in NLTE, where the uncertainty represents
the standard uncertainty of the mean, and an observed dispersion of
$\sigma_{A(\text{Na})} = 0.03 \, \text{dex}$ in LTE and 0.06\,dex in
NLTE. The small dispersion indicates that the Na abundances are
consistent within $1\sigma$, showing no significant star-to-star
variations. We estimated NLTE offsets from the INSPECT database%
\footnote{inspect-stars.com}. This database uses NLTE offsets from
\citet{lind_et_al2011}. 

Despite the absence of significant sodium abundance variations
among the four stars, this does not necessarily rule out a globular
cluster progenitor. First, our sample is small, and all four stars
may belong to the first-generation population, which typically
exhibits primordial Na abundances
\citep[e.g.,][]{2009carretta, 2019marino}.
A partly disrupted globular cluster, might have a stream that is
preferentially populated by first-generation stars, given that we expect
the first-generation stars to be preferentially lost first
\cite[e.\,g.][]{bekki_et_al2007mnras377_335B, d_ercole_et_al2008,
                decressin_et_al2008astronomy_and_astrophysics,
                decressin_et_al2010astronomy_and_astrophysics,
                2012gratton}.
For instance, \citet{vesperini_et_al2010_718l_112v} found in the results
of hydrodynamical simulations that the fraction of second-generation
stars in the Galactic halo is less than nine per cent. However, we also
note that a fully disrupted globular cluster might have both first- and
second-generation stars in its stream. Second, lower-mass or disrupted
clusters may host weaker abundance spreads, particularly if they formed
only a small second-generation population
\cite[e.g.,][]{usman_et_al2024}. Third, during tidal stripping, the
stars that populate the stream may not fully sample the abundance
distribution of the progenitor cluster. Therefore, the observed lack of
Na variation does not by itself exclude the globular cluster progenitor
scenario, especially in combination with the small metallicity
dispersion inferred from our MCMC analysis.

In summary, while our analysis cannot confirm that this tight clump
has a globular cluster origin, the analysis presented here cannot
rule out such an option either. We stress here again that the
spectra have quite low S/N to do a full abundance analysis and that
is why we stick to just the [Fe/H] from the CaT, which is from a
higher S/N region ($\sim$15-20), whereas the NaD region is already
a S/N of 7, and the S/N gets worse as we go down to blue regions
where we can find more useful lines.

\subsection{Constraints on the early phase of the infall of the 
            Sagittarius dwarf galaxy}\label{sec_traceback}
          
In order to investigate their nature and origin, we are interested 
where the cluster of four RGB stars would originally have been located
in the progenitor Sagittarius dwarf galaxy. To investigate this 
question, we selected stellar particles in the simulations by
\citet{vasiliev_et_al2021} and \citet{oria_et_al2022} to match the range
of possible 3D positions of the four RGB stars given the uncertainties
in their 3D positions (up to $\pm\! 5\sigma$). The resulting selection
of stellar particles in the final snapshot(s) were linked back to the
initial snapshot(s).  

\begin{figure*}
\includegraphics{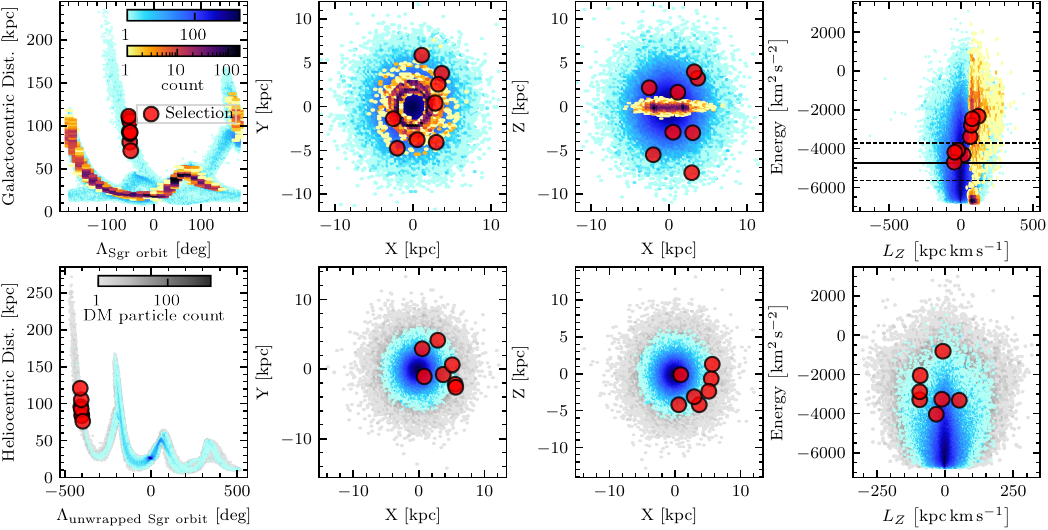}
\caption{Selection of stellar particles to match the data of the four
         RGB stars in the initial snapshots (except the left
         panels that show data of the final snapshots) of the
         simulations of the disruption of the Sagittarius dwarf galaxy
         in the time-dependent gravitational potential of both the
         Milky Way and Large Magellanic Cloud by
         \citet{vasiliev_et_al2021} (lower panels) and
         \citet{oria_et_al2022} (top panels). In all
         top panels the blueish background represents all stellar
         particles while the colored foreground are denoted as in the
         progenitor stellar disk in the simulation. In all lower panels
         the gray background and colored foreground are the dark matter
         and stellar particles, respectively. We use Cartesian
         positions in the middle panels and the right panels show the
         Cartesian Z-axis angular momentum. The interquartile range,
         including the median (solid, horizontal line) of the energies
         per mass of all stellar particles are demarcated by the
         dashed, horizontal lines in the top right panel.}
\label{sgr_selection_initial_final_figure}
\end{figure*}

Figure \ref{sgr_selection_initial_final_figure} shows energy and the
angular momentum $L _ Z$ across the Cartesian Z coordinate, for all
particles in the initial snapshots in the simulations by
\citet{vasiliev_et_al2021} and \citet{oria_et_al2022}. It is apparent
from the two top central panels that the selection of stellar particles
in the simulation by \citet{oria_et_al2022} is not associated to the
disk component but rather the halo in the progenitor galaxy. In both
simulations, the energies of the selection are all above the median
energy of the stellar particles (even if we take only those that are not
part of the disk component in the progenitor in the case of the
simulation by \citet{oria_et_al2022}). Statistically, they are thus less
bound to the progenitor than most of the other stellar particles, in
line with the fact that in these simulations
\cite[see also][]{s_l_j_gibbons_et_al2014} the stellar particles of
the Southern spur were lost early in the process of infall.

In the positional configuration of the initial snapshots, the selected
particles are also clustered at the edge of the distribution of the
stellar particles, which might indicate they once belonged to the halo
stellar population of the Sagittarius dwarf galaxy. Since the
simulations do not have a prescription for globular cluster/compact
stellar system formation and also not the necessary mass resolution
(1,000 solar masses in both simulations), we do not expect to find an
equivalent of a cluster in these simulations. Furthermore, these
simulations do not track chemical abundances. However we note that their
possible association with a halo population, would be in line with such
a scenario.

\subsection{Distant Southern versus Northern Sagittarius spurs}

\begin{figure*}
  \includegraphics{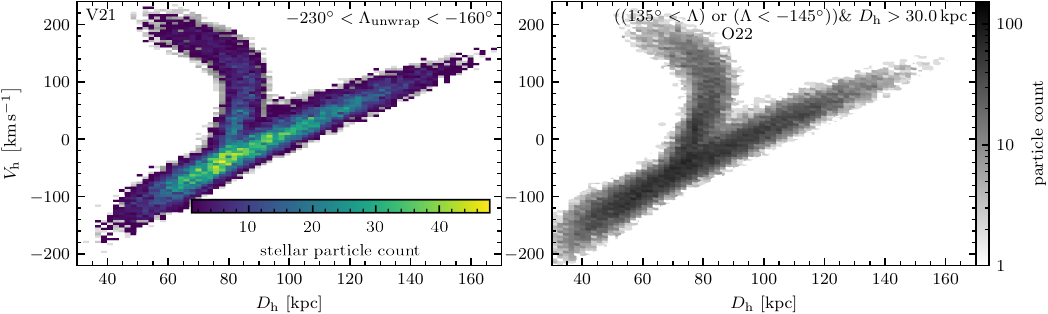}
  \caption{Line-of-sight velocity vs. heliocentric distance for the
           simulations by \citet{vasiliev_et_al2021} (V21, left
           column) and \citet{oria_et_al2022} (O22, right column) as in
           when particles are selected for the northern Sagittarius
           spur feature (see labels in the figure panels for details),
           color-coded by particle densities (with separate scales for
           the stellar and dark matter particles). These panels can be
           directly compared to the left and middle panels of Fig.
           \ref{sgrfar_data_figure} for the southern Sagittarius spur
           feature.}
  \label{northern_vs_southern_spur}
\end{figure*}

While RR Lyrae stars were found possibly in this feature in position
space by \citet{sesar_et_al2017b} and \citet{hernitschek_et_al2017} as
pointed out in the introduction, this is the first study to search for
the southern Sagittarius spur feature feature using full phase space
data and metallicities. Interestingly, there are several apparent
differences in the predictions for the distant, northern and southern
spurs in the simulations by \citet{vasiliev_et_al2021} and
\citet{oria_et_al2022} \cite[see also][]{dierickx_loeb2017}. First, we
see in \autoref{xz_plot} that the distant, northern spur is composed of
particles lost at two different times. One set was lost earlier between
2 - 3 billion years ago while the other set was lost more recent around
one billion year ago. The particles in distant, southern spur escaped
all at the earliest times during the infall between 2.5 - 3 billion
years ago. Second, both simulations predict the distant, northern spur
to be denser than the southern spur, indicating also that there is more
mass deposited in the northern spur than in the southern one (see lower
panels of the first and second column in \autoref{sgrfar_data_figure}
in comparison to the panels of \autoref{northern_vs_southern_spur}).
Third, there seem to a larger spread in the line-of-sight velocities of
the particles in the northern spur than in the southern spur. Finally,
the southern spur is predicted to go further out in terms of
heliocentric distance than the northern spur (see both panels of
\autoref{northern_vs_southern_spur}).

\subsection{Perspectives for future work}
The stellar data presented in this work is still limited in its
distance range. Given this limitation, and combined with the
sparsity of good quality data at these distances, we can not at
present make a clear distinction regarding the morphology of the
discovered overdensities. While we have been referring to the
feature as the southern spur of the Sagittarius stream, we can in
fact not clearly classify it morphologically and it could be also a
distant plume, or more generally a distant arm. More distant,
southern stars of the Sagittarius stream are needed to better
constrain the precise morphology. The picture that we have of this
part of the Sagittarius stream will hopefully change in the near
future with \textit{Gaia} DR4, the fifth-generation Sloan
Digital Sky Survey
\citep{kollmeier_et_al2026, vedant_chandra_et_al2026}, and
the Large Synoptic Survey Telescope \citep[LSST,][]{lsst}.

The stellar spectra presented in this work are of insufficient
signal-to-noise to measure other elements and provide a full
chemical characterisation of the clumped RGB, nor of the broader
Sagittarius southern spur candidates. Follow-up studies at higher
resolution and higher S/N ($> 30$) would be very beneficial for this
purpose. The addition of more stars to this clumped feature will -
together with their more detailed chemical analysis - also help
determine whether it has a globular cluster origin through a more
precise determination of the metallicity dispersion and (absence of)
the unique globular cluster signature of anti-correlation patterns
between chemical elements
\citep[e.g.,][]{2009carretta, 2012gratton}. The 4-metre Multi-%
Object Spectroscopic Telescope \citep{r_s_de_jong_et_al2019} S14
survey \citep{a_skuladottir_et_al2023} plans to observe the
Sagittarius stream and may provide helpful observations in this
part of the stream.
          
\section{Conclusions}\label{sec_conclusion}

In the search for close 5D comoving pairs of groups of stars in the
Milky Way halo we discovered a significant overdensity of RGB stars
that can be associated to the distant, southern Sagittarius stream
at heliocentric distances above 80~kpc. The additional line-of-sight
velocities from follow-up MIKE spectroscopy showed that the
four RGB stars cluster tightly in 6D space and agree with
predictions from the simulation of the disruption of the Sagittarius
stream in the evolving, gravitational potential of both the Milky
Way and Large Magellanic Cloud by \citet{vasiliev_et_al2021} for the
distant, southern spur. This distant, southern spur of the
Sagittarius stream is composed of stellar particles that were part
of the oldest debris lost during the early phases of the Sagittarius
dwarf galaxy infall 2.5 to 3 billion years ago.

The CaT-based metallicities for the four clustered RGB stars show a
low metallicity dispersion of $0.15 ^ {+0.17} _ {-0.08}$ around an
average [Fe/H] of] $-1.46 ^ {+0.11} _ {-0.09}$. Based on this
measurement, combined with the high clustering in 6D space, we
investigate that these four stars could be a remnant of a disrupted
Sagittarius globular cluster. Our measurements are however too
limited to fully confirm or refute this possibility. We note that
the simulation results of the Sagittarius stellar stream by
\citet{vasiliev_et_al2021} and \citet{oria_et_al2022} do suggest
that this debris is tracing the outer, less bound, stellar
components of the progenitor dwarf galaxy.

Despite the limited data currently available in the distant southern
Sagittarius stream, our analysis indicates that the compact cluster
of RGB stars may be part of a larger substructure within the
Sagittarius stream. This substructure is found to be coherent in 6D
phase space as derived from the value-added stellar distance catalog
SPDist of the DESI MWS DR 1.

This is the first spectroscopic study of the distant, southern
Sagittarius stream at heliocentric distances beyond 80~kpc, likely
composed mostly of the oldest debris of the Sagittarius stream. In
future work, these data can be used to constrain further the 
gravitational potential at large distances as well as the early
phase of the infall of the Sagittarius galaxy onto the Milky Way. 
\textit{Gaia} DR 4 and LSST will provide more data to better 
characterize the distant Galactic halo with anticipated larger
sample  sizes.

\section*{Acknowledgements}

  This paper includes data gathered with the 6.5 meter Magellan
  Telescopes located at Las Campanas Observatory, Chile. The authors
  like to thank the anonymous referee for a careful and constructive
  report that has certainly helped to improve the manuscript. We would
  like to thank Rohan Naidu for his contributions in the MIKE
  observations and discussions about the project. The authors would
  like to thank Amina~Helmi, Julio~Navarro, Sten~Sipma, and Martin~%
  Montelius for helpful comments that helped to improve this study.
  Moreover, the authors would like to cordially thank Eugene~%
  Vasiliev for providing the initial snapshot(s) of the simulation
  as part of \citet{vasiliev_et_al2021}. Finally, the authors would
  also like to thank Thomas~M.~Callingham for the help in handling
  isochrone data and giving a head start in working with Dark
  Energy Spectroscopic Instrument data. MB and ES acknowledge
  funding through VIDI grant `Pushing Galactic Archaeology to its
  limits' (with project number VI.Vidi.193.093) which is funded by
  the Dutch Research Council (NWO). This research has been
  partially funded from a Spinoza award by NWO (SPI 78-411). This
  research was supported by the International Space Science
  Institute (ISSI) in Bern, through ISSI International Team
  project 540 (The Early Milky Way).This work has received funding
  from the European Research Council (ERC) under the Horizon Europe
  research and innovation programme (Acronym: EARLYMW, Grant
  number: 101170507). AV gratefully acknowledges support from the 
  Canadian Institute for Theoretical Astrophysics (CITA) through a
  CITA National Fellowship and the International Astronomical Union
  (IAU) and the Gruber Foundation through a IAU Gruber Fellowship.
  APJ acknowledges support from the National Science Foundation
  under grant AST-2307599 and the Alfred P. Sloan Research
  Fellowship. GFT acknowledges support from the Agencia Estatal de
  Investigaci\'on del Ministerio de Ciencia en Innovaci\'on (AEI-MCIN)
  under grant number PID2023-150319NB-C21 and the grant RYC2024-051016-I
  funded by MCIN/AEI/10.13039/501100011033 and by the European Social
  Fund Plus. Co-funded by the European Union (Widening Participation,
  ExGal-Twin, GA 101158446). This work has made use of data from the
  European Space Agency (ESA) mission {\it Gaia} 
  (\url{https://www.cosmos.esa.int/gaia}), processed by the 
  {\it Gaia} Data Processing and Analysis Consortium (DPAC, 
  \url{https://www.cosmos.esa.int/web/gaia/dpac/consortium}). 
  Funding for the DPAC has been provided by national institutions,
  in particular the institutions participating in the {\it Gaia} 
  Multilateral Agreement. The Pan-STARRS1 Surveys (PS1) and the PS1
  public science archive have been made possible through
  contributions by the Institute for Astronomy, the University of
  Hawaii, the Pan-STARRS Project Office, the Max-Planck Society and
  its participating institutes, the Max Planck Institute for
  Astronomy, Heidelberg and the Max Planck Institute for
  Extraterrestrial Physics, Garching, The Johns Hopkins University,
  Durham University, the University of Edinburgh, the Queen's
  University Belfast, the Harvard-Smithsonian Center for 
  Astrophysics, the Las Cumbres Observatory Global Telescope Network
  Incorporated, the National Central University of Taiwan, the Space
  Telescope Science Institute, the National Aeronautics and Space
  Administration under Grant No. NNX08AR22G issued through the
  Planetary Science Division of the NASA Science Mission
  Directorate, the National Science Foundation Grant No. AST-%
  1238877, the University of Maryland, Eotvos Lorand University
  (ELTE), the Los Alamos National Laboratory, and the Gordon and
  Betty Moore Foundation. This research used data obtained with the
  Dark Energy Spectroscopic Instrument (DESI). DESI
  construction and operations is managed by the Lawrence Berkeley
  National Laboratory. This material is based upon work supported
  by the \href{https://www.energy.gov/}{U.S. Department of Energy},
  Office of Science, Office of High-Energy Physics, under Contract
  No. DE–AC02–05CH11231, and by the National Energy Research
  Scientific Computing Center, a DOE Office of Science User
  Facility under the same contract. Additional support for DESI was
  provided by the
  \href{https://www.nsf.gov/}{U.S. National Science Foundation}
  (NSF), Division of Astronomical Sciences under Contract No.
  AST-0950945 to the NSF’s National Optical-Infrared Astronomy
  Research Laboratory; the
  \href{https://stfc.ukri.org/}{Science and Technology Facilities 
                                Council of the United Kingdom};
  the \href{https://www.moore.org/}{Gordon and Betty Moore
                                    Foundation}; 
  the \href{https://www.hsfoundation.org/}{Heising-Simons
                                           Foundation}; 
  the \href{https://www.cea.fr/}{French Alternative Energies and
                                 Atomic Energy Commission}
  (CEA); the
  \href{https://secihti.mx/}{National Council of Humanities,
                             Science and Technology of Mexico}
  (CONAHCYT); the
  \href{http://www.mineco.gob.es/}{Ministry of Science and
                                   Innovation of Spain}
  (MICINN), and by the DESI Member Institutions: 
  \url{www.desi.lbl.gov/collaborating-institutions}. The DESI 
  collaboration is honored to be permitted to conduct scientific 
  research on I’oligam Du’ag (Kitt Peak), a mountain with
  particular significance to the
  \href{https://www.tonation-nsn.gov/}{Tohono O’odham Nation}. Any 
  opinions, findings, and conclusions or recommendations expressed
  in this material are those of the author(s) and do not
  necessarily reflect the views of the U.S. National Science
  Foundation, the U.S. Department of Energy, or any of the listed
  funding agencies. Funding for the Sloan Digital Sky Survey IV has
  been provided by the Alfred P. Sloan Foundation, the U.S.
  Department of Energy Office of Science, and the Participating
  Institutions. 

  SDSS-IV acknowledges support and resources from the Center for
  High Performance Computing  at the University of Utah. The SDSS
  website is www.sdss4.org.

  SDSS-IV is managed by the Astrophysical Research Consortium for
  the Participating Institutions of the SDSS Collaboration
  including the Brazilian Participation Group, the Carnegie
  Institution for Science,  Carnegie Mellon University, Center for
  Astrophysics | Harvard \& Smithsonian, the Chilean Participation
  Group, the French Participation Group, Instituto de Astrof\'isica
  de Canarias, The Johns Hopkins University, Kavli Institute for the
  Physics and Mathematics of the Universe (IPMU) / University of
  Tokyo, the Korean Participation Group, Lawrence Berkeley National
  Laboratory, Leibniz Institut f\"ur Astrophysik Potsdam (AIP),
  Max-Planck-Institut f\"ur Astronomie (MPIA Heidelberg),
  Max-Planck-Institut f\"ur Astrophysik (MPA Garching), 
  Max-Planck-Institut f\"ur Extraterrestrische Physik (MPE),
  National Astronomical Observatories of China, New Mexico State
  University, New York University, University of Notre Dame,
  Observat\'ario Nacional / MCTI, The Ohio State University,
  Pennsylvania State University, Shanghai Astronomical Observatory,
  United Kingdom Participation Group, Universidad Nacional
  Aut\'onoma de M\'exico, University of Arizona, University of
  Colorado Boulder, University of Oxford, University of Portsmouth,
  University of Utah, University of Virginia,
  University of Washington, University of Wisconsin, Vanderbilt
  University, and Yale University. The production of this work made use
  of the condor distributed computing software. We made use of the
  Action-based Galaxy Modeling Architecture code
  \citep[\texttt{AGAMA},][]{vasiliev2018, vasiliev2019a},
  \texttt{Astroquery} \citep{ginsburg_et_al2019},
  \texttt{Astropy} \citep{astropy2013, astropy2018, astropy2022},
  \texttt{colorcet} \cite[colorcet.com,][]{kovesi2015arxiv1509_03700},
  \texttt{corner} \citep{foreman-mackey2016}, \texttt{dustmaps} 
  \citep{green2018}, \texttt{emcee}
  \citep{foreman-mackey_et_al2013}, \texttt{Gala} \citep{gala},
  Image Reduction and Analysis Facility \citep{tody1986, tody1993, 
                                               iraf},
  \texttt{JupyterLab} \citep{kluyver_et_al2016}, \texttt{Kapteyn}
  \citep{terlouw_and_vogelaar2016}, \texttt{Korg} \citep{2023korg},
  \texttt{LESSPayne} \citep{Ji2025},
  \texttt{Matplotlib} \citep{hunter2007}, \texttt{NumPy}
  \citep{numpy}, \texttt{pandas}
  \citep{mckinney-proceeding-scipy-2010, reback2020pandas}, 
  \texttt{Polars}, \texttt{Scikit-learn} \citep{scikit-learn},
  \texttt{SciPy} \citep{scipy}, 
  \texttt{seaborn} \citep{waskom2021}, Starlink Tables
  Infrastructure Library Tool Set \citep{m_b_taylor2006}, Tool for
  OPerations on Catalogues And Tables \citep{m_b_taylor2005}, and
  \texttt{Vaex} \citep{breddels_veljanoski2018}. We would like to
  acknowledge the use of DeepL Write, an artificial intelligence
  writing assistant developed by DeepL, which was employed to
  enhance the readability and style of this article. We used
  Mistral Small 3 to edit the draft and sought guidance for
  developing Python code for the data analysis and plotting. Author
  contributions: All authors helped in finalising the paper draft.
  VC and APJ in MIKE data reduction to create the spectral dataset
  which is employed. Conceptualisation: MB, ES, AV, VC, APJ.
  Methodology: MB, ES, AV, VC, APJ. Software: MB, ES, AV, VC, APJ.
  Validation: MB, ES, AV, VC, APJ. Formal Analysis: MB, ES, AV, VC,
  APJ. Investigation: MB, ES, AV, VC, APJ. Data Curation: MB, ES, AV,
  VC, APJ. Writing - Original Draft: MB, ES, AV, VC, APJ. Writing -
  Review \& Editing: MB, ES, AV, VC, APJ, GFT. Visualisation: MB, ES,
  AV, VC, APJ, GFT.
  
\section*{Data Availability}
  The initial dataset that was used for the target selection is 
  available in the \textit{Gaia} and \textit{WISE} archives. The 
  reduced MIKE spectra of \textit{Gaia} (E)DR3 2598522907060164480, 
  2601287178076099712, 2601407505880573696, 2601408089995818752,
  and 2601835941752991872 are available upon reasonable request to
  the authors. The (third-party) literature samples include (i)
  simulation data from \citet{vasiliev_et_al2021} and
  \citet{oria_et_al2022} that is available at
  \url{zenodo.org/record/4038141} and 
  \url{people.ast.cam.ac.uk/~vasiliev/Sgr_init.npz}, and
  \url{doi.org/10.5281/zenodo.6581185}, respectively; (ii) blue 
  horizontal branch stars from the set of 
  \citet{manuel_bayer_et_al2025}; (iii) RR Lyrae 
  stars from \url{doi.org/10.26093/cds/vizier.51530204} and the
  online supplementary material of \citet{muraveva_et_al2025}; (iv)
  red giant stars from the DESI MWS archive; (v) candidate
  Sagittarius stream stars in the BOSS-MINESweeper catalog from the
  SDSS archive; and (vi) again data from the \textit{Gaia} archive.
  Additionally, the Pan-STARRS1 photometry for the sample of stars
  is available in the Pan-STARRS1 archive. Estimated total
  integrated interstellar dust reddening data along the line-of-%
  sight of each star from
  \citet{sfd, schlafly_finkbeiner2011} supporting the presented 
  photometry is available in \citet{sfd, schlafly_finkbeiner2011}. 
  Other data supporting this study are included within the tables
  of the article.

\bibliographystyle{mnras}
\bibliography{refs}

\appendix

\section{Sources of blue horizontal branch stars}%
\label{appendix_section_bhb_stars}
In our search for blue horizontal branch stars in existing catalogs 
that are in the same part of the halo of the Milky Way as the
cluster of four RGB stars in the distant, southern spur of the 
Sagittarius stream we checked \citet{xiang_x_xue_et_al2008},
\citet{ruhland_et_al2011}, \citet{xiang_x_xue_et_al2011}, 
\citet{guillaume_f_thomas_et_al2018}, \citet{vickers_et_al2021}, 
\citet{barbosa_et_al2022}, \citet{jie_ju_et_al2024}, 
and \citet{amarante_et_al2024} and did not find any due to 
missing coverage, and \citet{culpan_et_al2021} (distance coverage 
$\lesssim 60$~kpc), \citet{fengqing_yu_et_al2024} (up to
68.37~kpc), \citet{bystroem_et_al2025} did not find any due to
these catalogs not probing far enough.

\section{Candidate members of the distant, southern Sagittarius 
         stream in SPDist}%
\label{spdist_appendix_section}
         
We applied additional selections for the set of SPDist stars shown
in the third left panel of \autoref{sgrfar_data_figure} and the
selected candidates are listed in \autoref{spdist_table}.

\begin{table*}
  \caption{Stars in the value-added stellar distance catalog SPDist
           of the DESI MWS DR 1
           \citep{andrew_p_cooper_et_al2023, koposov_et_al2024,
                  koposov_et_al2026, desi_edr,
                  desi_dr1} 
           that we selected to be candidate members of a distant,
           southern feature of the Sagittarius stream to which we
           also associate a cluster of four RGB stars.}
  \begin{tabular}{llllllllllllll}
    \hline
    Source ID & R.A. & Dec. & $D _ \text{h}$
    & \multicolumn{2}{l}{$\mu ^ * _ \alpha$}
    & \multicolumn{2}{l}{$\mu _ \delta$}
    & \multicolumn{2}{l}{$V _ \text{h}$}
    & \multicolumn{2}{l}{[Fe/H]} & \\
    & \multicolumn{2}{l}{J2016} & [kpc]
    & \multicolumn{2}{l}{$\left[\text{mas}~\text{yr}^{-1}\right]$}
    & \multicolumn{2}{l}{$\left[\mathrm{mas\,yr^{-1}}\right]$}
    & \multicolumn{2}{l}{$\left[\text{km}~\text{s}^{-1}\right]$} & 
    &  \\
    \hline
    2436416199765387392 & 23h49m13s
    & $-7^\circ36{}^\prime13{}^{\prime\prime}$ 
    & $87 ^ {+12} _ {-12}$ & 0.3 & $\pm\! 0.2$ & $-0.2$ 
    & $\pm\! 0.2$ & $-165$ & $\!\pm 2$ & $-1.54$ & $\pm\! 0.07$ \\
    2541440279140626816 & 0h14m24s
    & $-2^\circ50{}^\prime37{}^{\prime\prime}$ 
    & $79 ^ {+12} _ {-12}$ & 0.7 & $\pm\! 0.4$ & $-0.3$
    & $\pm\! 0.2$ & $-188$ & $\!\pm 2$ & $-2.37$ & $\pm\! 0.08$ \\
    2604703257624429184 & 22h45m42s
    & $-11^\circ31{}^\prime48{}^{\prime\prime}$
    & $101 ^ {+15} _ {-15}$ & 0.2 & $\pm\! 0.2$ 
    & $-0.5$ & $\pm\! 0.2$ & $-218$ & $\!\pm 2$ & $-1.66$ 
    & $\pm\! 0.08$ \\
    2607762756103258496 & 22h45m51s
    & $-10^\circ58{}^\prime32{}^{\prime\prime}$ & $69 ^ {+6} _ {-6}$
    & 0.2 & $\pm\! 0.1$ & $-0.4$ & $\pm\! 0.1$ & $-163$ & $\!\pm 1$
    & $-1.25$ & $\pm\! 0.03$ \\
    2635601497165547008 & 23h10m03s
    & $-4^\circ31{}^\prime35{}^{\prime\prime}$
    & $64 ^ {+17} _ {-17}$ & 0.2 & $\pm\! 0.2$ & $-0.4$ 
    & $\pm\! 0.2$ & $-218$ & $\!\pm 2$ & $-1.62$ & $\pm\! 0.07$ \\
    2637025669665614336 & 23h17m59s
    & $-3^\circ47{}^\prime42{}^{\prime\prime}$
    & $92 ^ {+25} _ {-25}$ & 0.6 & $\pm\! 0.2$ & $-0.4$
    & $\pm\! 0.2$ & $-187$ & $\!\pm 2$ & $-2.32$ & $\pm\! 0.10$ \\
    2639188997448530816 & 23h36m50s
    & $-4^\circ16{}^\prime13{}^{\prime\prime}$ & $75 ^ {+9} _ {-9}$
    & 0.3 & $\pm\! 0.1$ & $-0.8$ & $\pm\! 0.1$ & $-186$ & $\!\pm 1$
    & $-1.60$ & $\pm\! 0.04$ \\
    2644888346035287936 & 23h23m23s
    & $-0^\circ18{}^\prime37{}^{\prime\prime}$ 
    & $72 ^ {+15} _ {-15}$& 0.3 & $\pm\! 0.2$ & $-0.7$
    & $\pm\! 0.2$ & $-217$ & $\!\pm 2$ & $-1.50$ & $\pm\! 0.09$ \\
    2738926212600476416 & 0h08m36s
    & $2^\circ19{}^\prime28{}^{\prime\prime}$ & $61 ^ {+6} _ {-6}$
    & 0.2 & $\pm\! 0.2$ & $-0.5$ & $\pm\! 0.1$ & $-232$ & $\!\pm 1$
    & $-2.42$ & $\pm\! 0.06$ \\
    2740069219361412736 & 0h04m51s
    & $3^\circ38{}^\prime20{}^{\prime\prime}$
    & $100 ^ {+20} _ {-20}$ & 0.5 & $\pm\! 0.3$ & $-0.3$
    & $\pm\! 0.2$ & $-237$ & $\!\pm 2$ & $-1.73$ & $\pm\! 0.09$ \\
    \hline
  \end{tabular}\\
  Source IDs, equatorial coordinates, and proper motions are from
  \textit{Gaia} DR 3 \citep{gaia_mission, gdr3}. The rest of the
  data is from the DESI MWS DR 1. The uncertainties of
  $D _ \text{h}$ were estimated following the recommended use of
  SPDist with regards to distance modulus and consequently distance
  calculations. The median of the absolute magnitude distribution in
  \textit{G} is taken, and the uncertainties for $M _ G$ are
  obtained from the $1\sigma$ quantiles (16th and 84th). The
  intrinsic precision of the method used to estimate the absolute
  magnitudes of stars in SPDist (0.167, 80\% relative distance
  precision) is added in quadrature. We report [Fe/H] from the
  pipeline called Stellar Parameters after the calibration
  described in section 4.2.1 in
  \citet{koposov_et_al2026}.
  \label{spdist_table}
\end{table*}

\bsp
\label{lastpage}
\end{document}